\documentclass[structabstract]{aa}  
\usepackage{graphicx}
\usepackage{txfonts}
\usepackage{natbib}
\usepackage{breqn}
\usepackage{longtable,lscape}
\def\teff{\mbox{$T_{\rm eff}$}}

\def\rv{\mbox{$R_{5495}$}}

\def\mum1{\mbox{$\mu$m$^{-1}$}}

\def\HII{\mbox{H\,{\sc ii}}}
\newcommand{\HeI}[1]{\mbox{He\,{\sc i}~$\lambda${#1}}}
\newcommand{\HeII}[1]{\mbox{He\,{\sc ii}~$\lambda${#1}}}
\newcommand{\CIII}[1]{\mbox{C\,{\sc iii}~$\lambda${#1}}}
\newcommand{\NIII}[1]{\mbox{N\,{\sc iii}~$\lambda${#1}}}

\newcommand{\SiIII}[1]{\mbox{Si\,{\sc iii}~$\lambda${#1}}}

\begin{document}

   \title{A search for Galactic runaway stars using \\ {\it Gaia} Data Release 1 and {\it Hipparcos} proper motions}
%   \subtitle{II. The foreground ISM}

%   \subtitle{I. Overviewing the $\kappa$-mechanism}

   \author{J. Ma{\'\i}z Apell{\'a}niz\inst{1}
           \and
           M. Pantaleoni Gonz\'alez\inst{1,2}
           \and
           R. H. Barb\'a\inst{3}
           \and
           S. Sim\'on-D{\'\i}az\inst{4,5}
           \and
           I. Negueruela\inst{6}
           \and
           D. J. Lennon\inst{7}
           \and
           A.~Sota\inst{8}
           \and
           E. Trigueros P\'aez\inst{1,6}
%                 \and
%                 N. R. Walborn\inst{4}
%                 \and
%                 A. Pellerin\inst{5}
%                  \and
%                 A. Sota\inst{3}
%                  \and
%                 S. Sim\'on-D{\'\i}az\inst{4,5}
%                 \and
%                 A. Marco\inst{2}
%                 \and
%                 J. Alonso-Santiago\inst{2}
%                 \and
%                 J. Sanchez Bermudez\inst{8}
%                 \and
%                 R. C. Gamen\inst{9}
%                 \and
%                 J. Lorenzo\inst{2}
%                  \and
%                  others
          }

   \institute{Centro de Astrobiolog{\'\i}a, CSIC-INTA. Campus ESAC. Camino bajo del castillo s/n. E-28\,692 Villanueva de la Ca\~nada. Spain. \\
              \email{jmaiz@cab.inta-csic.es} \\
         \and
              Departamento de Astrof{\'\i}sica y F{\'\i}sica de la Atm\'osfera. Universidad Complutense de Madrid. E-28\,040 Madrid. Spain. \\
         \and
              Departamento de F{\'\i}sica y Astronom{\'\i}a. Universidad de La Serena. Av. Cisternas 1200 Norte. La Serena. Chile. \\
         \and
              Instituto de Astrof{\'\i}sica de Canarias. E-38\,200 La Laguna. Tenerife. Spain. \\
         \and
              Departamento de Astrof{\'\i}sica. Universidad de La Laguna. E-38\,205 La Laguna. Tenerife. Spain. \\
         \and
              Departamento de F{\'\i}sica. Ingenier{\'\i}a de Sistemas y Teor{\'\i}a de la Se\~nal. Escuela Polit\'ecnica Superior. Universidad de Alicante. Carretera San Vicente del Raspeig s/n. E-03\,690 San Vicente del Raspeig. Spain. \\
         \and
              ESA. European Space Astronomy Centre. Camino bajo del castillo s/n. E-28\,692 Villanueva de la Ca\~nada. Spain. \\
         \and
              Instituto de Astrof{\'\i}sica de Andaluc{\'\i}a-CSIC. Glorieta de la Astronom\'{\i}a s/n. E-18\,008 Granada. Spain. \\
%         \and
%              Space Telescope Science Institute, 3700 San Martin Drive, Baltimore, MD 21\,218, USA \\
%         \and
%              Department of Physics \& Astronomy, 1 College Circle, SUNY Geneseo, Geneseo, NY 14\,454, USA \\
%         \and
%              Instituto de Astrof\'{\i}sica de La Plata (CCT La Plata-CONICET, Universidad Nacional de La Plata), Paseo del Bosque s/n, 1900 La Plata, Argentina \\
             }

   \date{Received 8 February 2018; accepted 20 April 2018}

% \abstract{}{}{}{}{} 
% 5 {} token are mandatory
 
  \abstract
  {The first Gaia Data Release (DR1) significantly improved the previously available proper motions for the majority of the Tycho-2 stars.}
  {We want to detect runaway stars using Gaia DR1 proper motions and compare our results with previous searches.}
  {Runaway O stars and BA supergiants are detected using a 2-D proper-motion method. The sample is selected using Simbad, spectra from our GOSSS project,
  literature spectral types, and photometry processed using CHORIZOS.}
  {We detect 76 runaway stars, 17 (possibly 19) of them with no prior identification as such, with an estimated detection rate of approximately one half of 
   the real runaway fraction. An age effect appears to be present, with objects of spectral subtype B1 and later having travelled for longer distances than 
   runaways of earlier subtypes. We also tentatively propose that the fraction of runaways is lower among BA supergiants that among O stars but further 
   studies using future Gaia data releases are needed to confirm this. The frequency of fast rotators is high among runaway O stars, which indicates that a
   significant fraction of them (and possibly a majority) is produced in supernova explosions.}
  {}


   \maketitle
%
%________________________________________________________________

\section{Introduction}

$\,\!$ \indent Runaway stars are (usually massive) Population I stars that move at large peculiar velocities with respect 
to the mean Galactic rotation. Two mechanisms were proposed over half-a-century ago to explain their production: 
the ejection of a close companion in a supernova explosion \citep{Blaa61} and a three- or more-body interaction
at the core of a compact stellar cluster \citep{Poveetal67}. Runaway stars can be detected by their proper motions, 
radial velocities, or a combination of both. The availability of good-quality proper motions from {\it Hipparcos}
allowed many new runaway stars to be detected \citep{Hoogetal01,Mdzi04,MdziChar05,Tetzetal11}.

The success in the detection of runaway stars with {\it Hipparcos} led to great expectations for {\it Gaia}.
On 14 September 2016 the first {\it Gaia} Data Release (DR1) was presented \citep{Browetal16}. 
Gaia DR1 includes parallaxes and proper motions from TGAS (Tycho-Gaia Astrometric Solution, \citealt{Michetal15}) 
for the majority (but not all) of the Tycho-2 stars. Among the excluded Tycho-2 stars we find all of the very bright 
objects but also some dimmer ones. TGAS proper motions exist for a significantly larger number of stars
than for {\it Hipparcos} and, for the stars in common between both catalogs, they are more precise.

Having good-quality proper motions and/or radial velocities is a requirement to detect runaway 
stars\footnote{Even though other characteristics such as the presence of a bow shock can provide hints of their
nature.} but we also need to correctly identify a sample of massive stars to do so, as such objects are a needle in a 
haystack of hot evolved low-mass stars % CHANGE
where the range of proper motions and radial velocities can be very large due to their
relative proximity and mixture of populations.
In this respect, the Galactic O-Star Spectroscopic Survey (GOSSS, \citealt{Maizetal11}) can play a fundamental role.
GOSSS is obtaining $R\sim$2500, high-S/N, blue-violet spectroscopy of all optically accessible Galactic O stars.
To this date, three survey papers \citep{Sotaetal11a,Sotaetal14,Maizetal16} have been published
with a total of 590 O stars\footnote{The GOSSS spectra are being gathered with six facilities: 
the 1.5~m Telescope at the Observatorio de Sierra Nevada (OSN); 
the 2.5~m du Pont Telescope at Las Campanas Observatory (LCO); 
the 3.5~m Telescope at the Observatorio de Calar Alto (CAHA); 
and the 2.0~m Liverpool Telescope (LT),
the 4.2~m William Herschel Telescope (WHT), 
and the 10.4~m Gran Telescopio Canarias (GTC) at the Observatorio del Roque 
de los Muchachos (ORM). Of those, the LT is a recent addition to the mix}. 
Several additional hundreds of O stars and several thousands of B and later-type stars 
have already been observed and their data will be published in the near future. Why is GOSSS needed? One reason is
given by \citet{Maizetal13}: at the time that paper was published, 24.9\% of the O stars with previous spectral 
classifications GOSSS had observed turned out to be of other spectral type (some were even late-type stars, 
\citealt{Maizetal16}). Five years later, the number of such false positives in GOSSS is $\sim$35\%. Therefore, 
finding an object with a peculiar proper motion that Simbad says is an O star could be a new runaway star or it could
be something else: one needs some confirmation such as a good spectrogram before being certain.

A preliminary version of the results in this paper was presented in \citet{Maizetal17d}, from now on Paper I. That
contribution was limited to O stars. Here we include additional O stars in our sample, extend it to BA supergiants, 
present new spectra and information, and discuss our procedures and results in more detail, as Paper I was a 
relatively brief non-refereed contribution to conference proceedings. 

\section{Data and methods}

$\,\!$ \indent The core data of this paper are the {\it Gaia} DR1 and {\it Hipparcos} proper motions for Galactic 
massive stars presented in the next subsection. We also use supporting data in the form of spectra, photometry, and
images, which are presented in the following subsections.

\subsection{Proper motions}

$\,\!$ \indent When {\it Gaia} DR1 was announced, our initial plan was to analyze the included TGAS parallaxes to 
increase the meager number of useful trigonometric distances available for O stars \citep{vanL07a,Maizetal08a}. 
However, the TGAS parallaxes for O stars provide little new information, as the brightest O stars are not included 
and only one star, AE~Aur, has $\pi_{\rm o}/\sigma_\pi >$ 6, where $\pi_{\rm o}$ is the observed parallax and 
$\sigma_\pi$ is the parallax uncertainty. It should be remembered that, in general,
$<\!d\!>\; \ne 1/\pi_{\rm o}$, that is, the inverse of the observed parallax is not an unbiased estimator of 
the trigonometric distance \citep{LutzKelk73,Maiz01a,Maiz05c}.

On the other hand, the TGAS proper motions proved to have useful information. We used 
three % four 
different samples:

\begin{enumerate}
 \item O stars already observed with GOSSS when this work started. This was the sample in Paper I.
 \item Additional objects for which an O-type classification appears in Simbad.
% \item Objects classified as WR stars in Simbad.
 \item Objects classified as BA supergiants in Simbad.
\end{enumerate}

The first sample is a
``clean'' sample: uniformly selected, with a coherent magnitude limit over the whole sky (complete to $B$ = 8 and near-complete
to $B$ = 10), and with few expected false positives\footnote{We have found that ALS~18\,929, which we had previously claimed as 
O9.7, is actually a B0 after obtaining better data, see below. There are two other similar cases that we plan to publish soon and 
two other stars in GOSSS-III that we suspect may actually be sdO. Of those five objects, only ALS~18\,929 has Hipparcos or TGAS 
proper motions and is therefore relevant for this paper.}. 
%The third sample, WR stars, is also clean, as WR stars are relatively easy to identify and an 
%excellent-quality catalog exists \citep{vadH06}. WR stars may even become the first stellar class for which a near-complete 
%catalog for the whole Milky Way becomes available \citep{Sharetal12a}. 
The situation is different for the other two samples, 
which are ``dirtier''. Therefore, after cross-matching them
%samples 2 and 4 
with entries with TGAS and/or Hipparcos proper motions we had to clean them, as explained below.

\begin{table}
\caption{Samples used in this paper. T and H refer to TGAS and Hipparcos, respectively, and those columns indicate the number
of entries in those catalogs. R is the number of runaway candidates, R\% is the percentage of runaways, and D is the number 
of discarded objects.}
\label{samples}
\begin{center}
\begin{tabular}{llrrrlrr}
\hline
Sample  & Sp. ty. & T    & H    & T or H & \,R    & R\% & D  \\
\hline
 1+2    & O       &  770 &  325 &  871   & 48$^1$ & 5.7 & 26 \\ % 29 + 12 (no 13) + 7, 48/(871-26)
%3      & WR      &   69 &   67 &   79   &  3     & 4.0 &  4 \\
 3      & BA I    &  833 &  427 &  959   & 27     & 3.1 & 79 \\ % 27/(959-79)
\hline
\end{tabular}
\end{center}
$^1$ Not including ALS~18\,929, see below.
\end{table}

We cross-matched each of the samples with the TGAS and Hipparcos \citep{vanL07a} catalogs. The numbers of objects found are
given in Table~\ref{samples}. Note that since some objects have both TGAS and Hipparcos proper motions the third number in each
row is always less than the sum of the other two (bright objects are generally absent from TGAS and dim ones from Hipparcos). 
When an object has entries in both catalogs, we choose TGAS. For all samples we did an initial cleaning by excluding all cases
where the uncertainties in the proper motions were too large to be useful for a determination of the runaway character of the
object. For the last two samples
%samples 2 and 4 
we did an additional cleaning for the stars where the analysis indicated the possibility of them
being runaways.

\begin{itemize}
 \item For sample 2 we first attempted to obtain a GOSSS spectrum (subsection 2.2). This confirmed the nature of some of the 
       objects as O stars and discarded others. For the rest we searched the literature for photometry and conducted a CHORIZOS 
       analysis to estimate their \teff\ (subsection 3.3). This also confirmed some as O stars and discarded others. Finally,
       those objects without GOSSS spectral types or a CHORIZOS analysis were also discarded.
 \item For sample 3
%4 
       we also first attempted to obtain a GOSSS spectrum, which confirmed the nature of part of the sample and
       discarded another part. For the rest we searched the literature for spectral classifications and retained only those
       coming from references with a proven record of coincidence with GOSSS spectral types (see \citealt{Maizetal04b} and 
       GOSSS-I for discussions on this issue). The rest were discarded. We did not attempt a CHORIZOS analysis for BA 
       supergiants, as the main contaminant of the sample are BA stars of lower luminosity, which are difficult to
       distinguish from photometry alone. 
\end{itemize}

Finally, we also excluded some objects that are too close to us, as the method used to find runaway stars can yield false positives
in that case, though we retained some close known runaways such as $\zeta$~Oph.
The total number of discarded objects for one reason or another is given in Table~\ref{samples}. Clearly, sample 3
%4 
is the dirtiest of the three and we found the worst offenders to be the alleged A supergiants: Simbad contains many objects with such
classifications that turn out to be something else. Our strategy is designed to minimize the number of false positives at the risk of 
increasing the number 
of % CHANGE
false negatives because the latter will likely be found in the incoming years using future {\it Gaia} data releases.
In other words, in order to eliminate (or at least largely reduce) our noise we are forced to reduce our signal. 

Having established how we selected our sample, we now describe how we detected the runaway-star candidates. % CHANGE
Ideally, to identify runaway stars one computes the 3-D motions of the stars and compares them to the expected velocities at
their location using a Galactic rotation model. That requires knowledge not only of the proper motions but also of the
distances and radial velocities. Since we do not have precise measurements of the latter two for most of the sample, we are
forced to follow a modification of the 2-D proper-motion-only method used by \citet{Moffetal98} to identify WR and O-star 
runaways with Hipparcos. 

\begin{figure*}
\centerline{\includegraphics[width=\linewidth]{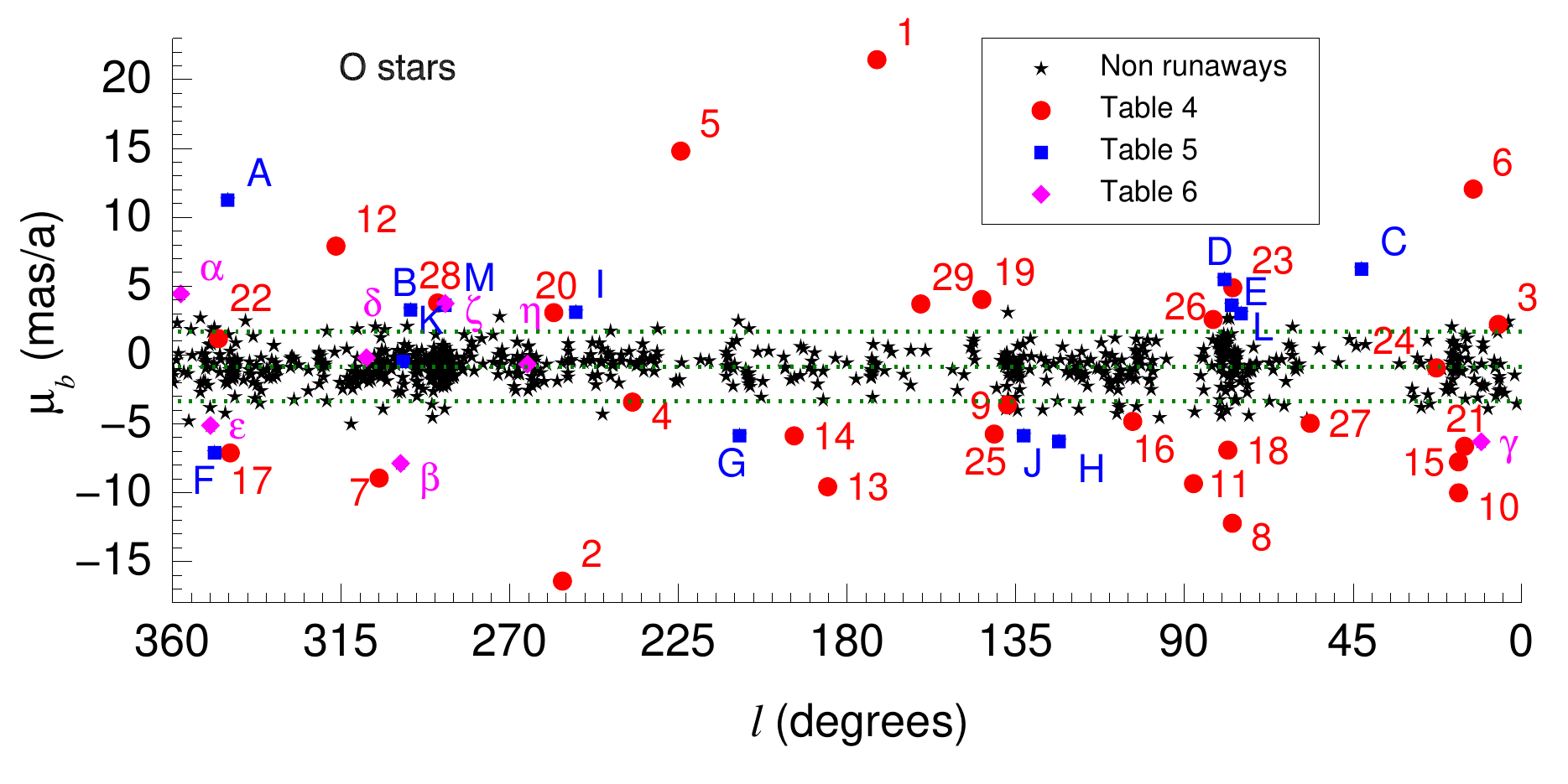}}
\centerline{\includegraphics[width=\linewidth]{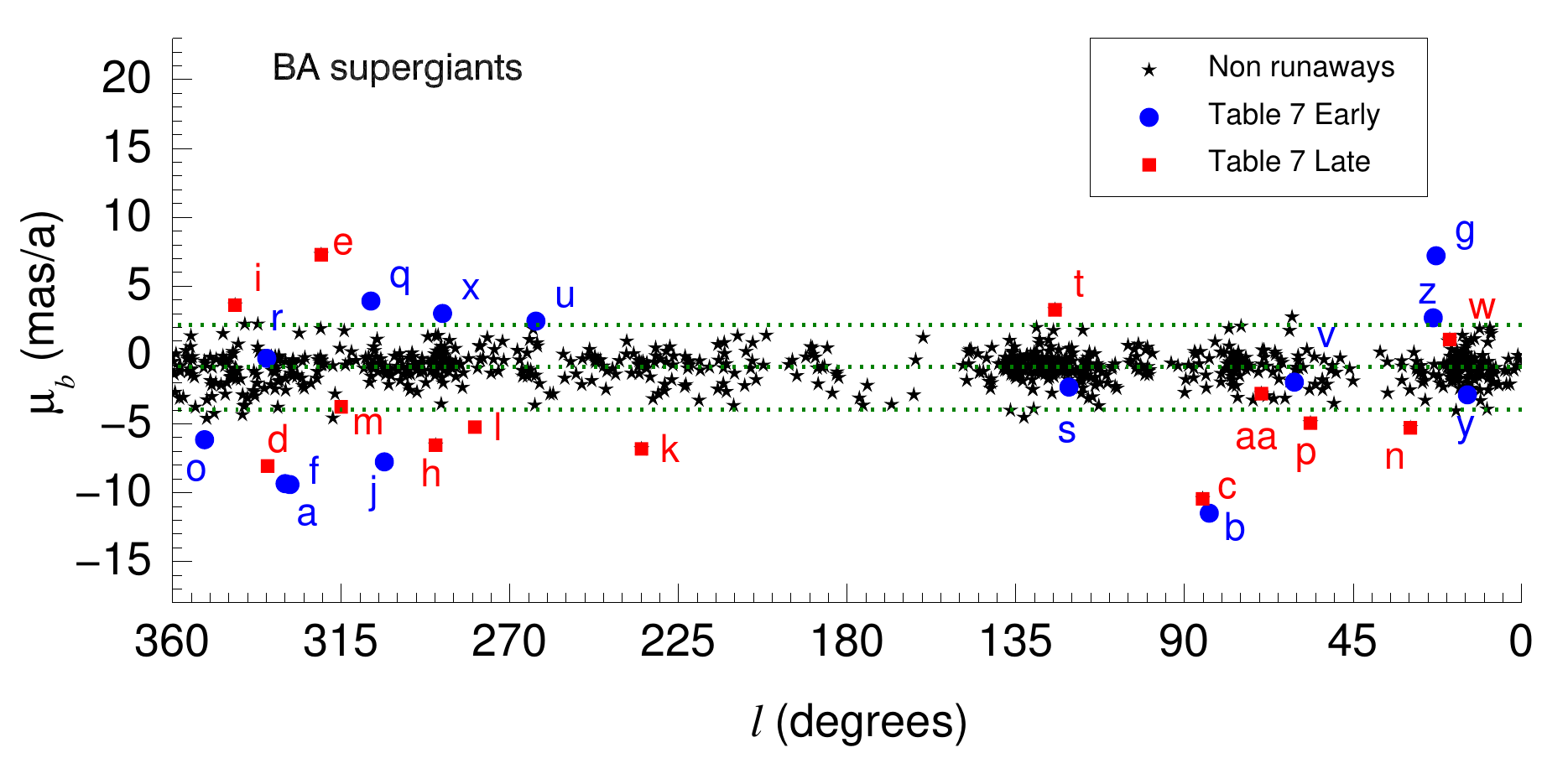}}
\caption{Observed proper motion in Galactic latitude for the O stars (top panel) and the BA supergiants (bottom panel) in this paper.
Different colors and symbols are used to identify the runaways candidates in Tables~\ref{tabs1},~\ref{tabs2},~\ref{tabs3a}~and~\ref{tabs3b},
with the IDs in those tables also shown. The dotted green lines represent the functions and 2$\sigma$ deviations used to detect runaway candidates.
Note that the vertical scales are the same in both panels to allow for an easier comparison.}
\label{pmlat}
\end{figure*}

\begin{figure*}
\centerline{\includegraphics[width=\linewidth]{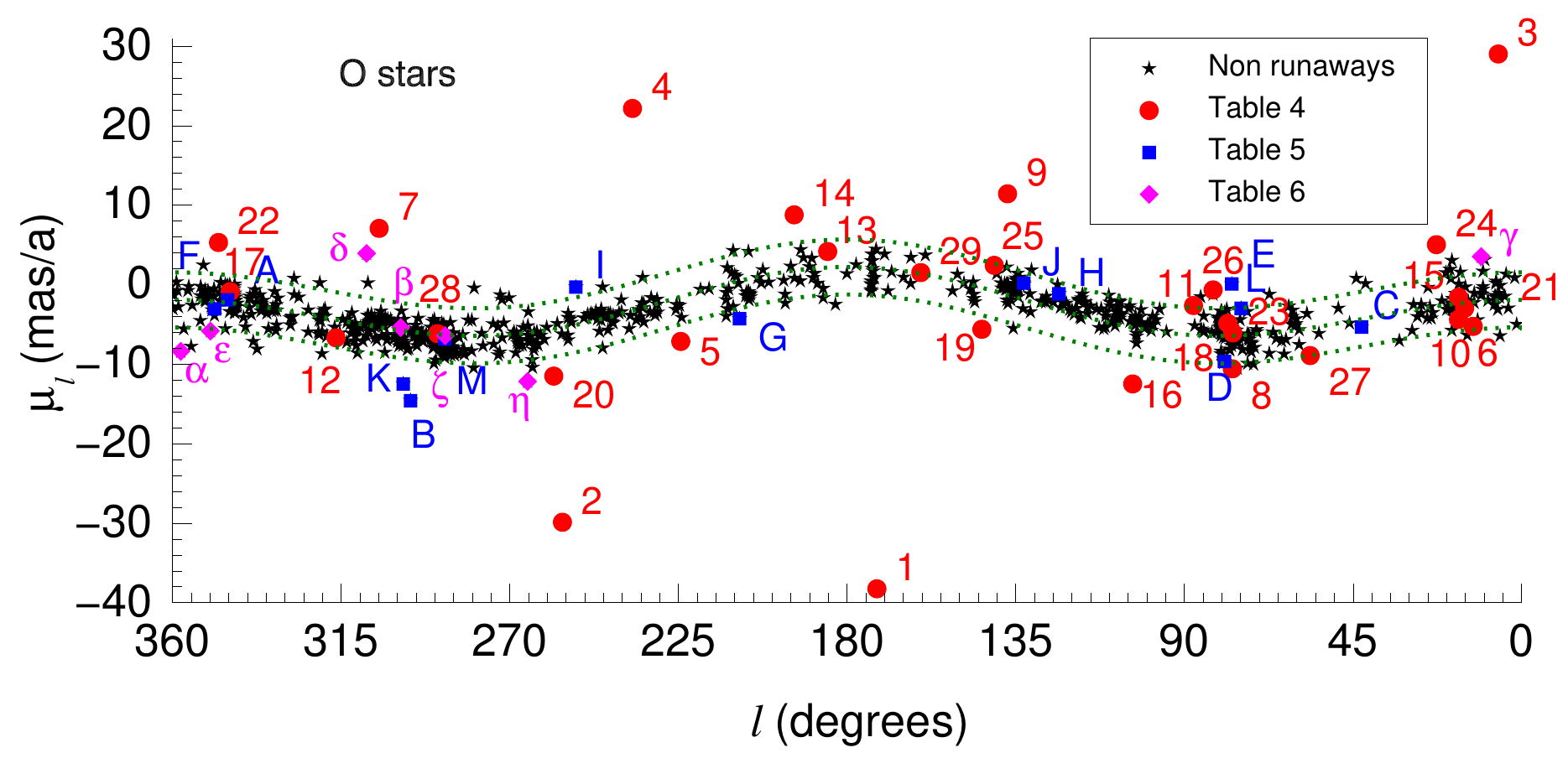}}
\centerline{\includegraphics[width=\linewidth]{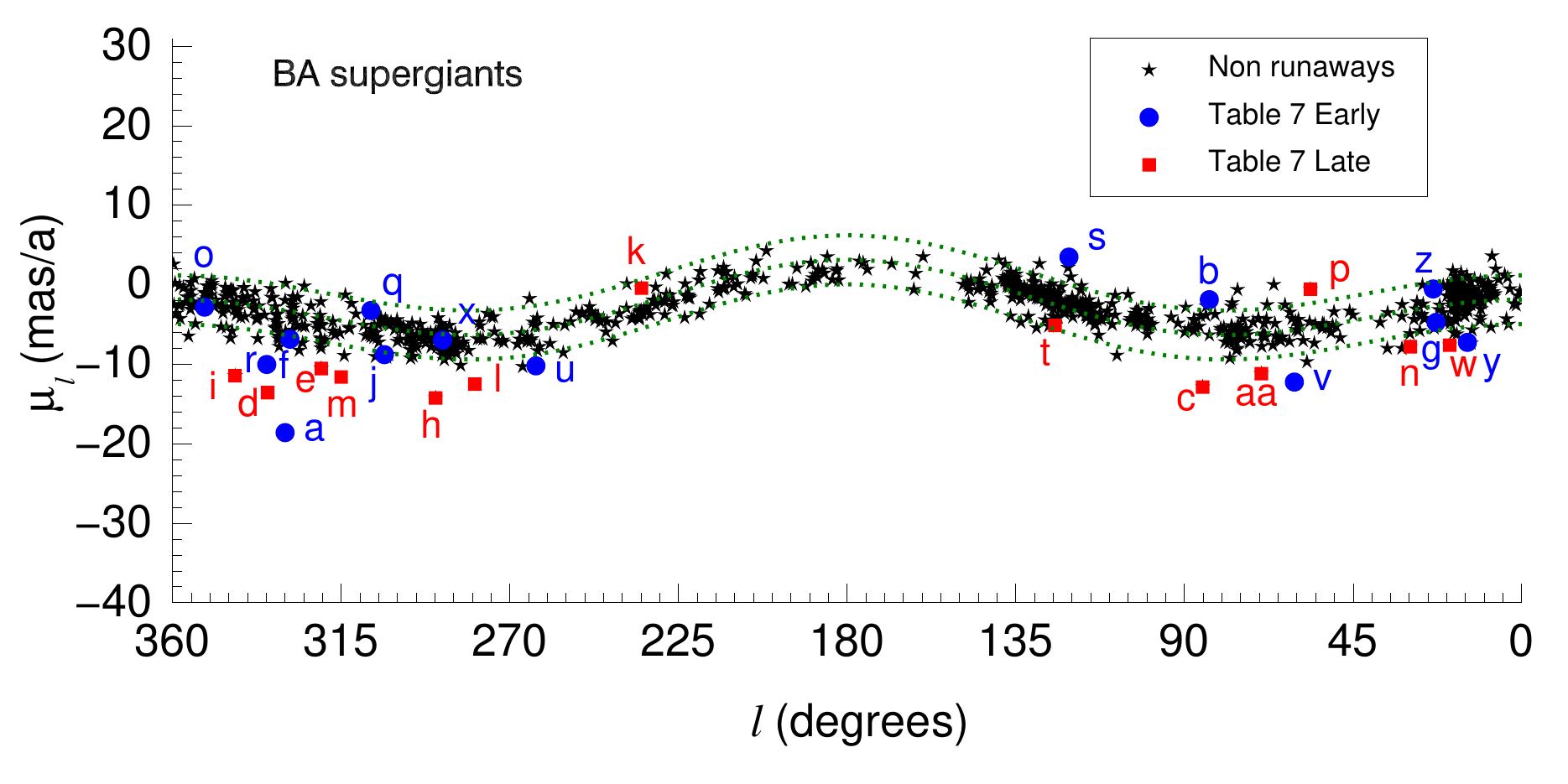}}
\caption{Observed proper motion in Galactic longitude for the O stars (top panel) and the BA supergiants (bottom panel) in this paper.
Different colors and symbols are used to identify the runaways candidates in Tables~\ref{tabs1},~\ref{tabs2},~\ref{tabs3a}~and~\ref{tabs3b},
with the IDs in those tables also shown. The dotted green lines represent the functions and 2$\sigma$ deviations used to detect runaway candidates.
Note that the vertical scales are the same in both panels to allow for an easier comparison.}
\label{pmlon}
\end{figure*}

For all of the objects in our samples, % CHANGE
the proper motions in RA ($\mu_\alpha$) and declination ($\mu_\delta$) were transformed into their 
equivalents in Galactic latitude ($\mu_b$) and longitude ($\mu_l$). 
Then, % CHANGE
a robust mean for $\mu_b$ (reflecting
the solar motion in the vertical direction), $<\!\mu_b\!>$, and a robust standard deviation,
$\sigma_{\mu_b}$, were calculated
both for the O stars (samples 1+2) and the BA supergiants (sample 3). % CHANGE
For $\mu_l$ we robustly fitted a functional form  
$f(l) = a_0 + a_1\cos l + a_2\cos 2l$ and we also calculated the robust standard deviation, $\sigma_{\mu_l}$,
from the fit. Results are shown in 
Table~\ref{parameters} % CHANGE
and Figs.~\ref{pmlat}~and~\ref{pmlon}. 

To detect runaway stars we computed the normalized difference (in standard deviations) of the difference between 
the observed proper motions and the fitted ones i.e.:

\begin{displaymath}
\Delta = \sqrt{\left(\frac{\mu_b^\prime}{\sigma_{\mu_b}}\right)^2 + \left(\frac{\mu_l^\prime}{\sigma_{\mu_l}}\right)^2},
\end{displaymath}

\noindent where $\mu_b^\prime = \mu_b-<\!\mu_b\!>$ and $\mu_l^\prime = \mu_l-f(l)$ are the corrected proper motions, 
and sorted the results from largest to smallest. The cut in $\Delta$ is the same in both cases and was empirically 
established at 3.5 by comparing our results with those of the 3-D method of \citet{Tetzetal11},
who use a threshold of 28~km/s for the peculiar velocity of a runaway star. % CHANGE
This 2-D method is simpler than a full computation of the 3-D velocities and
has the advantage of being self-contained and, therefore, less prone to errors introduced by
the required external measurements in the 3-D method (distances and radial velocities). 
However, it can yield false positives and negatives, which we analyze later on.

\subsection{GOSSS spectra}

\begin{table*}
\caption{New GOSSS spectral classifications sorted by Galactic longitude. GOS/GAS/GBS stands for Galactic O/B/A Star. The last column 
indicates which GOSSS 
paper (if any) was the star included in. The information in this table is also available in electronic form at the GOSC web site
({\tt http://gosc.cab.inta-csic.es}) and at the CDS via anonymous ftp to {\tt cdsarc.u-strasbg.fr} (130.79.128.5) or via 
{\tt http://cdsweb.u-strasbg.fr/cgi-bin/qcat?J/A+A/}.}
\label{tabGOSSS}
\begin{center}
\begin{tabular}{lcccllllc}
\hline
Name              & GOSSS ID               & R.A. (J2000) & Decl. (J2000)  & ST    & LC  & Qual.    & Second .   & Paper  \\
\hline
ALS 4962          & GOS 010.85$+$03.27\_01 & 18:21:46.166 & $-$21:06:04.42 & ON5   & I   & fp       & \ldots     & \ldots \\
HD 171\,012       & GBS 014.52$-$04.38\_01 & 18:33:10.096 & $-$18:22:06.18 & B0.2  & Ia  & \ldots   & \ldots     & \ldots \\
BD $-08$ 4617     & GOS 022.79$+$00.99\_01 & 18:29:14.330 & $-$08:33:40.91 & O8.5  & III & (n)      & \ldots     & \ldots \\
BD $-08$ 4623     & GBS 022.88$+$00.66\_01 & 18:30:34.797 & $-$08:38:03.69 & B0.5: & Ia: & \ldots   & \ldots     & \ldots \\
HD 161\,961       & GBS 023.64$+$12.90\_01 & 17:48:36.856 & $-$02:11:46.29 & B0.5  & Ib  & \ldots   & \ldots     & \ldots \\
67 Oph            & GBS 029.73$+$12.63\_01 & 18:00:38.716 & $+$02:55.53.63 & B5    & Ib  & \ldots   & \ldots     & \ldots \\
ALS 18\,929       & GBS 042.79$+$10.57\_01 & 18:31:01.379 & $+$13:30:12.85 & B0    & V   & \ldots   & \ldots     & III    \\
HD 161\,695       & GAS 056.40$+$26.94\_01 & 17:45:40.235 & $+$31:30:16.84 & A0    & Ib  & \ldots   & \ldots     & \ldots \\
HD 190\,066       & GBS 060.69$-$04.54\_01 & 20:02:22.102 & $+$22:09:05.26 & B0.7  & Iab & \ldots   & \ldots     & \ldots \\
%Tyc 3159-00006-1  & GOS 078.83$+$03.15\_01 & 20:18:40.368 & $+$41:32:45.01 & O9.7  & Iab & \ldots   & \ldots     & \ldots \\
ALS 11\,244       & GOS 079.36$+$02.61\_01 & 20:22:37.775 & $+$41:40:29.15 & O4.5  & III & (n)(fc)p & \ldots     & \ldots \\
69 Cyg            & GBS 083.39$-$09.96\_01 & 21:25:47.025 & $+$36:40:02.59 & B0.2  & Iab & \ldots   & \ldots     & \ldots \\
HD 215\,733       & GBS 085.16$-$36.35\_01 & 22:47:02.508 & $+$17:13:59.00 & B1    & Ib  & \ldots   & \ldots     & \ldots \\
$\kappa$ Cas      & GBS 120.84$+$00.14\_01 & 00:32:59.987 & $+$62:55:54.43 & BC0.7 & Ia  & \ldots   & \ldots     & \ldots \\
HD 8065           & GAS 124.57$+$15.96\_01 & 01:23:45.801 & $+$78:43:33.81 & A0    & Iab & \ldots   & \ldots     & \ldots \\
$\rho$ Leo AB     & GBS 234.89$+$52.77\_01 & 10:32:48.671 & $+$09:18:23.71 & B1    & Ib  & Nstr     & \ldots     & \ldots \\
CPD $-$34 2135    & GOS 252.40$-$00.04\_01 & 08:13:35.361 & $-$34:28:43.93 & O7.5  & Ib  & (f)p     & \ldots     & \ldots \\
GP Vel            & GBS 263.06$+$03.93\_01 & 09:02:06.860 & $-$40:33:16.90 & B0.5  & Ia  & \ldots   & \ldots     & \ldots \\
HD 94\,909        & GBS 287.96$+$01.94\_01 & 10:56:24.465 & $-$55:33:04.85 & B0    & Ia  & \ldots   & \ldots     & \ldots \\
HD 86\,606        & GBS 289.82$-$13.13\_01 & 09:56:09.735 & $-$71:23:21.49 & B1    & Ib  & \ldots   & \ldots     & \ldots \\
AB Cru            & GOS 298.47$+$04.41\_01 & 12:17:37.122 & $-$58:09:52.44 & O8    & III & \ldots   & BN0.2: Ib: & \ldots \\
HD 115\,842       & GBS 307.08$+$06.83\_01 & 13:20:48.339 & $-$55:48:02.49 & B0.5  & Ia  & \ldots   & \ldots     & \ldots \\
HD 156\,359       & GBS 328.68$-$14.52\_01 & 17:21:18.725 & $-$62:55:05.36 & B0    & Ia  & \ldots   & \ldots     & \ldots \\
HD 150\,898       & GBS 329.98$-$08.47\_01 & 16:47:19.657 & $-$58:20:29.19 & B0    & Ib  & \ldots   & \ldots     & \ldots \\
CPD $-$50 9557    & GBS 334.87$-$02.73\_01 & 16:39:07.728 & $-$50:54:08.29 & B0    & Ib  & \ldots   & \ldots     & \ldots \\
HD 167\,756       & GBS 351.47$-$12.30\_01 & 18:18:40.156 & $-$42:17:18.22 & B0.2  & Ib  & \ldots   & \ldots     & \ldots \\
\hline
\end{tabular}
\end{center}
\end{table*}

\begin{figure*}
\centerline{\includegraphics[width=\linewidth]{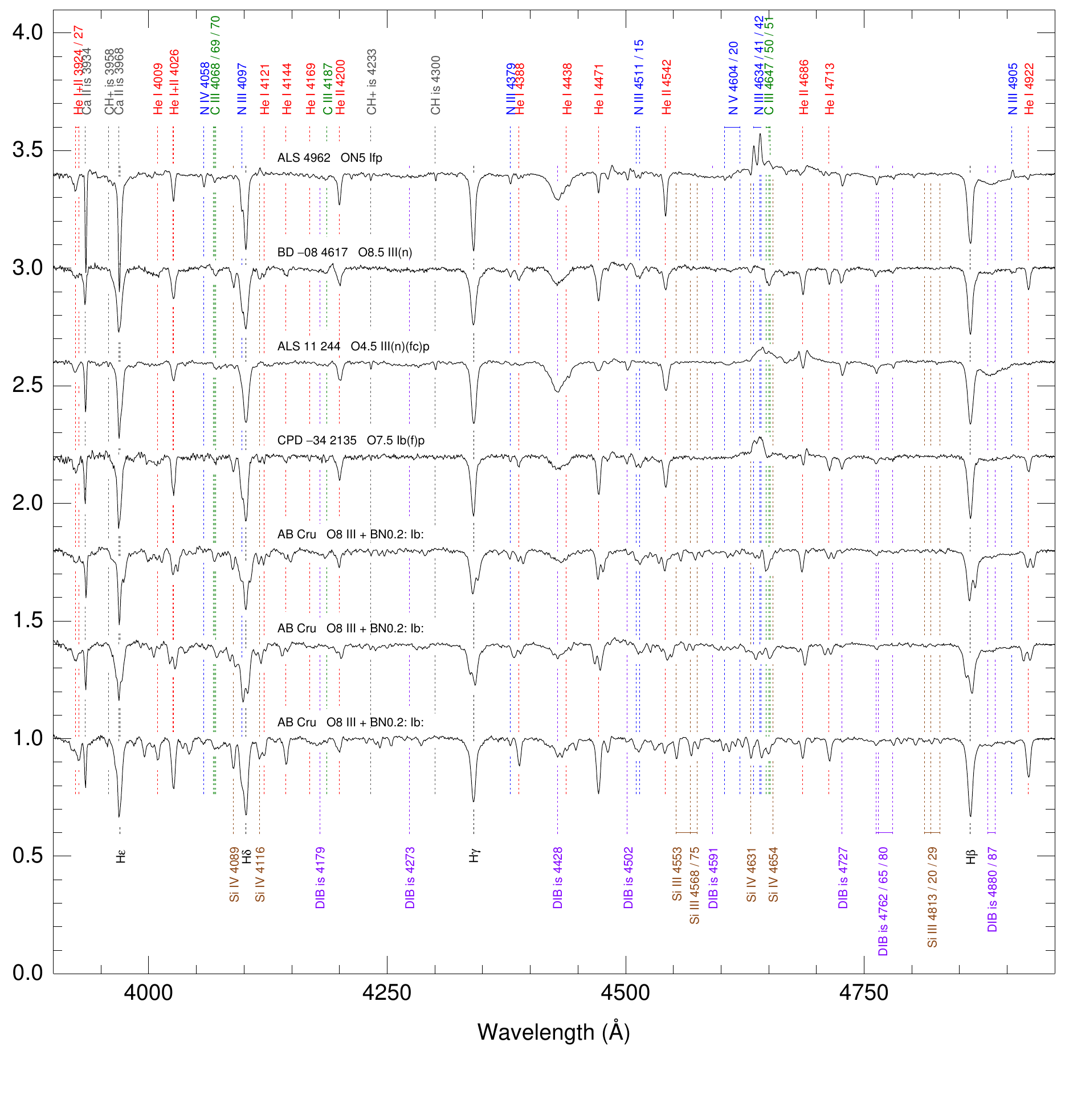}}
\caption{GOSSS rectified spectrograms for the O stars in Table~\ref{tabGOSSS} sorted by Galactic longitude. For the SB2 
AB~Cru three different orbital phases are shown.}
\label{GOSSS_O}
\end{figure*}

\begin{figure*}
\centerline{\includegraphics[width=\linewidth]{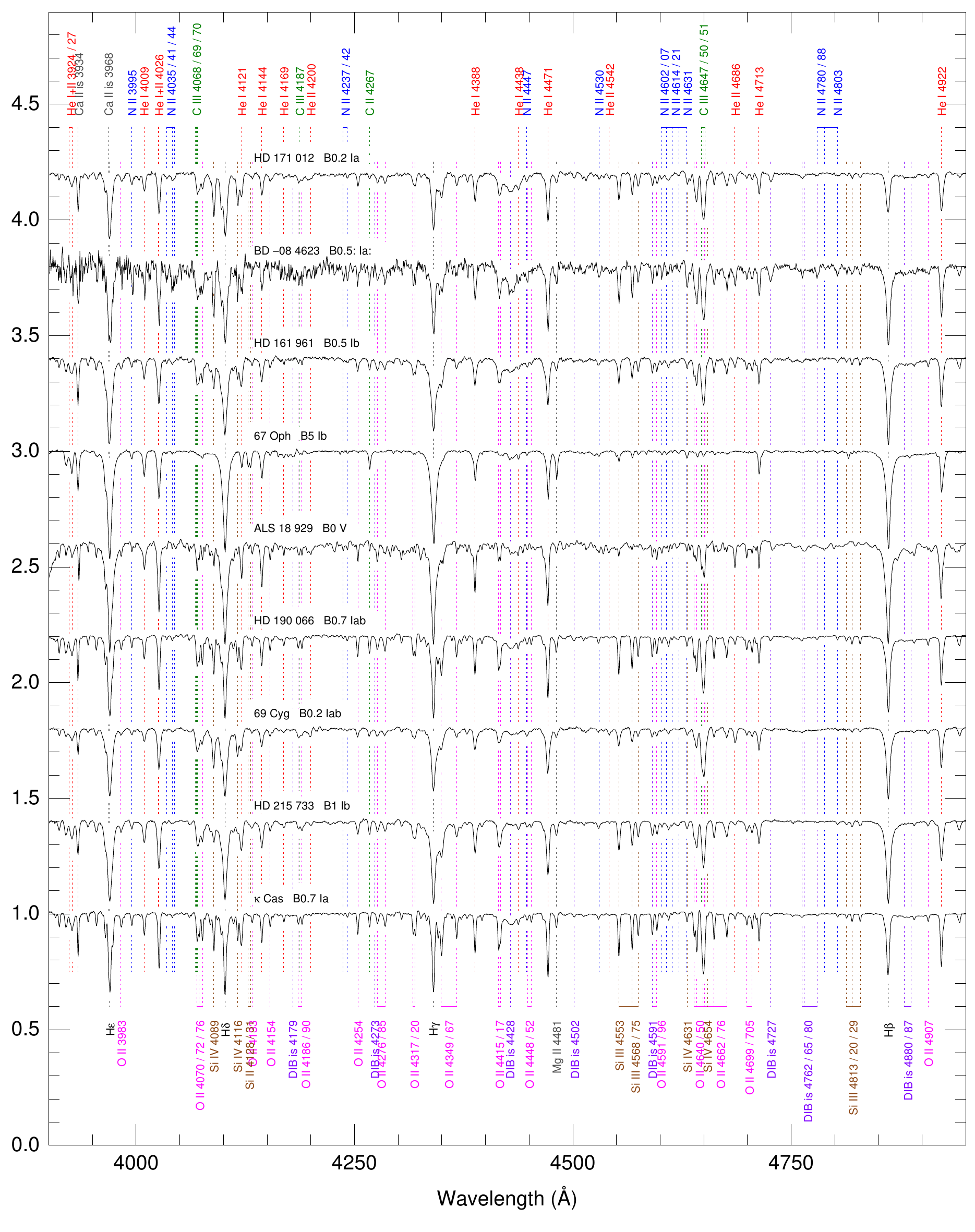}}
\caption{GOSSS rectified spectrograms for the B stars in Table~\ref{tabGOSSS} sorted by Galactic longitude.}
\label{GOSSS_B}
\end{figure*}

\addtocounter{figure}{-1}

\begin{figure*}
\centerline{\includegraphics[width=\linewidth]{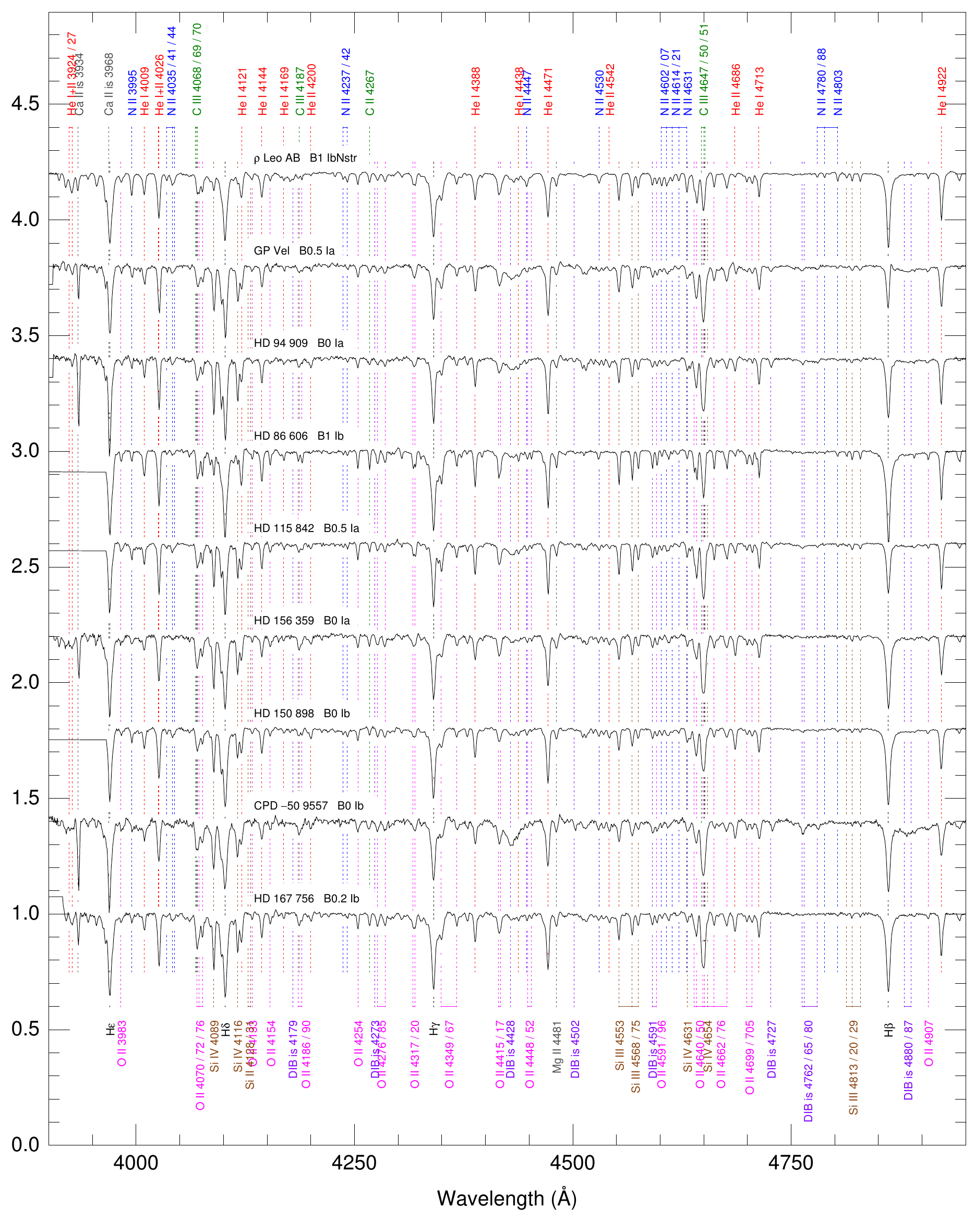}}
\caption{(continued)}
\end{figure*}

\begin{figure*}
\centerline{\includegraphics[width=\linewidth]{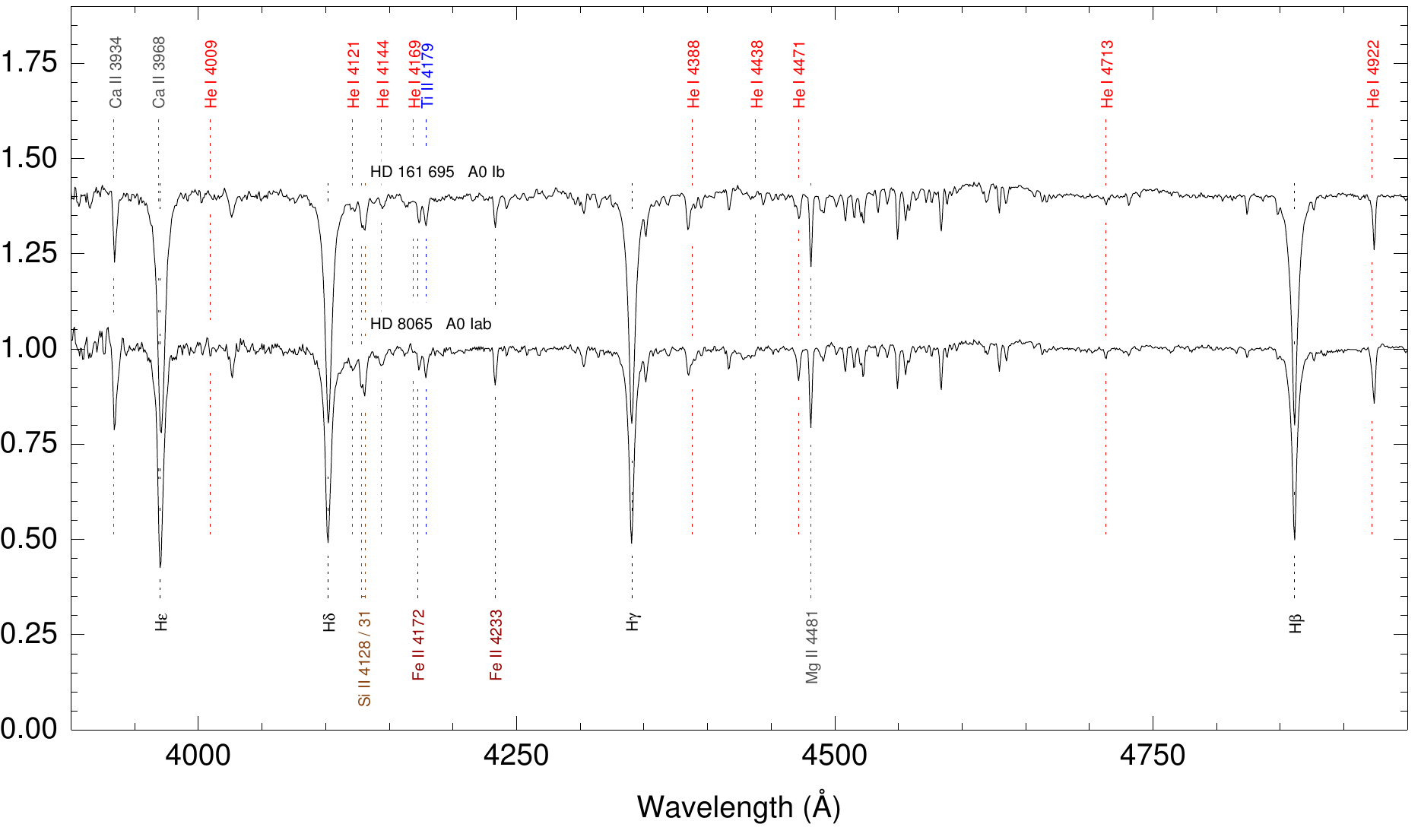}}
\caption{GOSSS rectified spectrograms for the A stars in Table~\ref{tabGOSSS} sorted by Galactic longitude.}
\label{GOSSS_A}
\end{figure*}

$\,\!$\indent To confirm the spectral type of the runaway candidates, one should ideally obtain good-quality 
spectroscopy. That is what we did for the 25 objects in Table~\ref{tabGOSSS} as part of the GOSSS project. The spectrograms are
shown in Figs.~\ref{GOSSS_O},~\ref{GOSSS_B},~and~\ref{GOSSS_A}.
The spectral classification was done using MGB 
\citep{Maizetal12,Maizetal15b} and a new grid of spectroscopic standards that extends to A0 for luminosity classes II to V and to
A7 for supergiants 
(Ma{\'\i}z Apell\'aniz et al., in preparation). The new spectrograms are presented in 
Figs.~\ref{GOSSS_O},~\ref{GOSSS_B}~and~\ref{GOSSS_A}
and the new spectral types are given in Table~\ref{tabGOSSS}. Each of the new stars is discussed in the Results section.
The addition of the new objects leaves the GOSSS statistics with a total of 594 % 590 + 5 - 1
O stars, 20 other early-type stars, and 11
late-type stars. The spectrograms and spectral types are available from the Galactic O-Star Catalog (GOSC,
\citealt{Maizetal04a,Maizetal17c}) web site. GOSC has been recently moved to a new URL ({\tt http://gosc.cab.inta-csic.es})
but the old one ({\tt http://gosc.iaa.es}) will be kept as a mirror, at least temporarily.

\subsection{Photometry and CHORIZOS}

$\,\!$\indent Unfortunately, we were not able to obtain good-quality spectroscopy for all of our runaway candidates
in time for the publication of this paper. For the stars without spectral types we collected photometry from the literature 
and used CHORIZOS \citep{Maiz04c} to estimate their effective temperatures. For details on how this is done, we refer the 
reader to \citet{Maizetal14a}, where we explain how this procedure can be carried on without significant biases, and to 
\citet{MaizBarb18}, where we use photometry similar to the one in this paper to measure 
extinction for a large sample of Galactic stars (see \citealt{Ariaetal06,Maizetal07,Maizetal15a,SimDetal15a,Damietal17a} for 
further examples). In particular, for this paper:

\begin{itemize}
 \item We did not include in the analysis those objects without accurate photometry at both sides of the Balmer jump
       (Johnson $UB$ or Str\"omgren $uv$), as for them is not possible to adequately measure the effective
       temperature ($T_{\rm eff}$) of OBA stars.
 \item We used the Milky Way grid of \citet{Maiz13a} and the family of extinction laws of \citet{Maizetal14a}.
 \item For each star we left $T_{\rm eff}$, $E(4405-5495)$ (amount of extinction), and $R_{5495}$ (type of extinction) as
       free parameters, and $\log d$ (logarithmic distance) as a dummy parameter. The luminosity class (LC) was fixed.
\end{itemize}

\begin{table*}
 \caption{Results of the CHORIZOS runs. The photometry (Phot.) column gives the bands included in each run with the same nomenclature as
         in Ma{\'\i}z Apell\'aniz \& Barb\'a (submitted to A\&A) i.e. J for Johnson $UBV$, 2 for 2MASS $JHK_{\rm S}$, G for 
         Gaia $G$, T for Tycho-2 $BV$, and S for Str\"omgren $uvby$.}
\label{chorizos}
\begin{center}
\renewcommand{\arraystretch}{1.4}
\begin{tabular}{lcccccccl}
\hline
Name              & R.A. (J2000) & Decl. (J2000)  & Phot. & $T_{\rm eff}$ (kK)   & LC  & $E(4405-5495)$            & $R_{5495}$             & Conclusion                 \\
\hline
CPD $-57$ 3781    & 10:46:11.507 & $-$58:39:12.39 & J2GT  & $39.3^{+7.1}_{-6.9}$ & 5.0 & $0.825^{+0.019}_{-0.025}$ & $3.64^{+0.12}_{-0.11}$ & Likely O                   \\
%CPD $-59$ 3300    & 11:17:07.618 & $-$60:22:39.93 & J2GT  & $45.5^{+3.5}_{-5.0}$ & 5.0 & $0.800^{+0.019}_{-0.025}$ & $3.74^{+0.12}_{-0.12}$ & Likely O                   \\
CPD $-72$ 1184    & 11:58:59.966 & $-$73:25:48.37 & J2TS  & $32.4^{+5.9}_{-2.6}$ & 5.0 & $0.198^{+0.035}_{-0.037}$ & $3.66^{+0.72}_{-0.54}$ & Likely O, possibly early B \\
CPD $-63$ 2886    & 13:39:35.390 & $-$63:39:41.74 & J2GT  & $42.0^{+5.6}_{-6.9}$ & 5.0 & $1.306^{+0.025}_{-0.025}$ & $3.16^{+0.06}_{-0.08}$ & Likely O                   \\
%CPD $-60$ 5274    & 14:17:47.262 & $-$61:30:50.90 & J2G   & $21.4^{+1.7}_{-1.5}$ & 5.0 & $0.544^{+0.031}_{-0.031}$ & $3.72^{+0.20}_{-0.20}$ & Likely B                   \\
HDE 322\,987      & 17:18:39.102 & $-$37:20:16.79 & J2GT  & $41.1^{+6.1}_{-6.8}$ & 5.0 & $1.200^{+0.019}_{-0.025}$ & $3.24^{+0.08}_{-0.06}$ & Likely O                   \\
HDE 315\,927      & 17:31:57.315 & $-$29:38:36.96 & J2GTS & $39.5^{+6.8}_{-5.5}$ & 5.0 & $1.156^{+0.013}_{-0.019}$ & $3.04^{+0.06}_{-0.05}$ & Likely O                   \\
\hline
\end{tabular}
\renewcommand{\arraystretch}{1.0}
\end{center}
\end{table*}

The results of our CHORIZOS analysis are shown in Table~\ref{chorizos} and are discussed in the next section. We emphasize that
the identifications as O stars based on photometry should be taken with care until a spectroscopic confirmation is obtained.

\subsection{WISE imaging}

\begin{figure*}
\centerline{\includegraphics[width=0.85\linewidth]{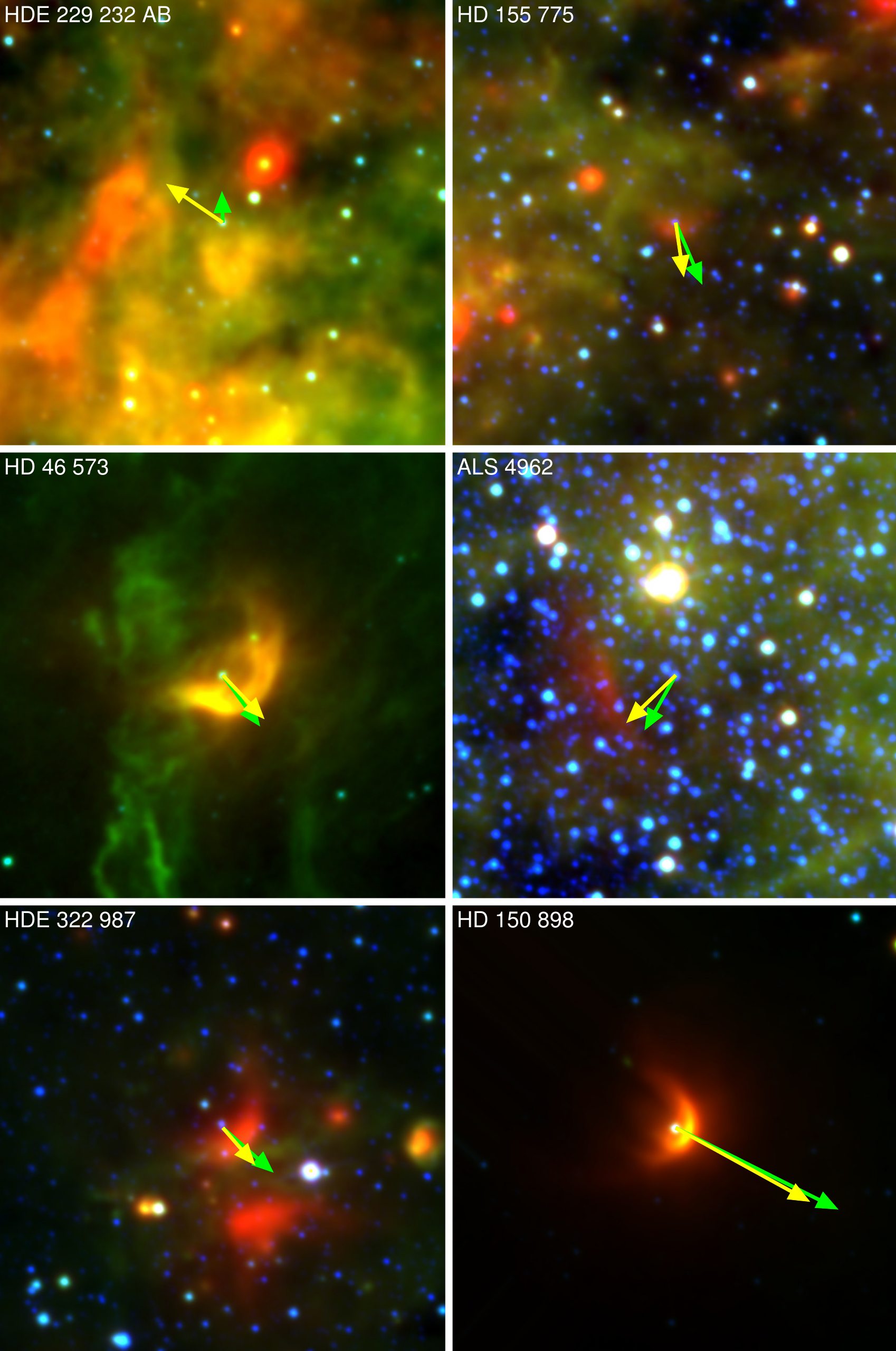}}
\caption{WISE W4+W3+W2 RGB mosaics for twelve of the runaway candidates in this paper sorted by their order of appearance. Each field is 
$14\arcmin\times14\arcmin$ and is oriented with Galactic (not equatorial) North towards the top and Galactic East towards the left. In each mosaic the
 runaway candidate is at the center and the arrows show the original proper motion (green) and the corrected one (yellow).}
\label{WISE}
\end{figure*}

\addtocounter{figure}{-1}

\begin{figure*}
\centerline{\includegraphics[width=0.85\linewidth]{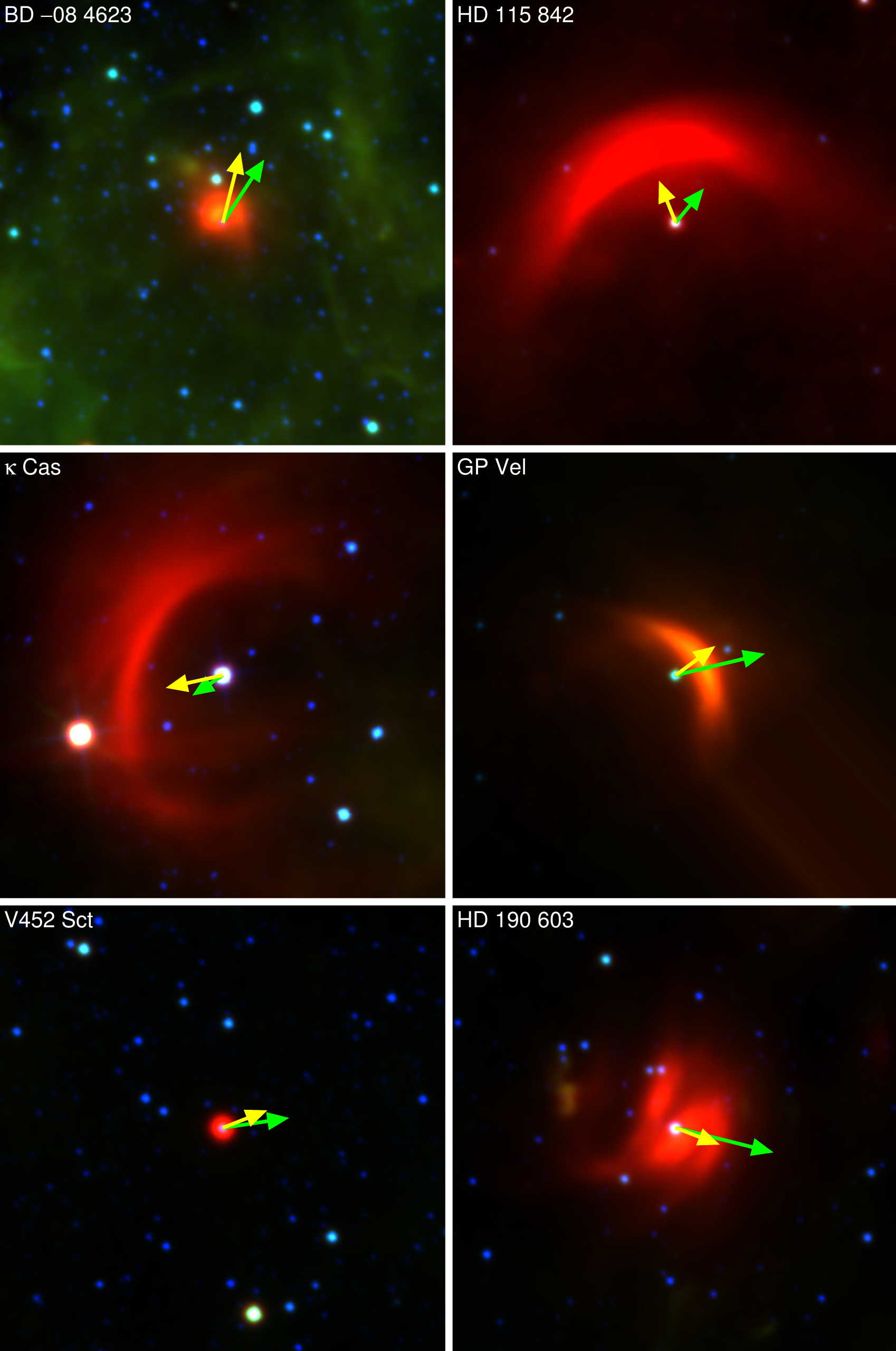}}
\caption{(continued)}
\end{figure*}

$\,\!$\indent For each of the new runaway candidates we searched for the dust emission from possible bow shocks in the {\it WISE} W3 and W4 images. In most
cases we did not find any sign of one. However, in some of them there are bow shocks or other interesting structures that we show in Fig.~\ref{WISE} as
RGB mosaics using the W4+W3+W2 channels.

\section{Runaway candidates}

$\,\!$\indent In this section we provide information about each of the new objects that we have identified as possible runaways.
Given the differences in our previous knowledge about the runaway character of the objects in 
this paper, for the sake of clarity we divide them in three categories: [a] runaways known before Paper I, [b]
objects described as runaways for the first time in paper I, and [c] runaway candidates not in Paper I. Each category is presented
in the next three subsections.

\subsection{Previously known O-type runaways in Paper I}

\begin{table*}
\caption{Previously known O-type runaway stars from Table~1 in Paper~1. The list is sorted
by $\Delta$, the normalized deviation from the mean latitude and longitude proper motions for their 
 Galactic longitude. An ID is provided for each star in order to identify it in Figs.~\ref{pmlat}~and~\ref{pmlon}. The T/H flag 
indicates the origin of the proper motions (TGAS or Hipparcos, respectively). Some relevant references
are listed in the last column. 
See Tables~\ref{tabs2},~\ref{tabs3a},~and~\ref{tabs3b} for the rest of the runaways in this paper.} % CHANGE
\label{tabs1}
\begin{center}
\begin{tabular}{rlrrcl}
\hline
ID & Name              & $\mu_b^\prime$ & $\mu_l^\prime$  & T/H & Ref.             \\
\hline
 1 & AE Aur            &    22.29       & $-$40.33        & T   & B54,B01,H01,L12  \\ % & -5.0 &  1.0 & -5.0 &  1.0 &             V
 2 & $\zeta$ Pup       & $-$15.58       & $-$24.49        & H   & C74,N97,H01,S08  \\ % & -5.0 &  1.0 & -5.0 &  1.0 & (n)  I
 3 & $\zeta$ Oph       &     3.05       &    31.00        & H   & B61,G79,H01,T10  \\ % & -5.0 &  1.0 & -5.0 &  1.0 & nn   IV
 4 & $\mu$ Col         &  $-$2.59       &    25.43        & H   & B54,B01,H01,T11  \\ % & -5.0 &  1.0 & -5.0 & -2.0 &             V
 5 & HD 57\,682        &    15.64       &  $-$5.63        & T   & M04,G12,S14,P15  \\ % & -5.0 & -3.0 & -5.0 &  1.0 &             IV
 6 & HD 157\,857       &    12.89       &  $-$3.05        & H   & B61,M04,S08,T11  \\ % & -1.0 & -5.0 & -5.0 &  1.0 &             II
 7 & HD 116\,852       &  $-$8.11       &    12.43        & H   & C74,M98,M04,S08  \\ % & -5.0 &  1.0 &  8.0 & -2.0 &             II-III
 8 & Y Cyg             & $-$11.38       &  $-$4.17        & T   & G85,M05a,S08,T11 \\ % & -5.0 & -5.0 & -5.0 & -2.0 &             IV
 9 & HD 17\,520 A      &  $-$2.78       &    12.72        & H   & M04,D12,S14      \\ % & -5.0 &  1.0 & 10.0 & -1.0 &             V
10 & V479 Sct          &  $-$9.17       &  $-$2.00        & T   & R02,M12,M16      \\ % & +8.0 & -6.0 & -5.0 & -2.0 &             V
11 & 68 Cyg            &  $-$8.51       &     3.71        & H   & B61,W89,S08,T11  \\ % & 10.0 &  1.0 & -4.0 & -1.0 & n    III
12 & HD 124\,979       &     8.74       &  $-$2.27        & T   & T11,D12,S14      \\ % & 10.0 & -7.0 & -5.0 &  1.0 & (n)  IV
13 & HD 36\,879        &  $-$8.74       &     2.01        & T   & M04,M05,M16      \\ % & -2.0 &  1.0 & -5.0 & -1.0 & (n)  V
14 & HD 41\,997        &  $-$5.03       &     7.00        & T   & C74,M98,M04,S08  \\ % & -5.0 &  1.0 & -5.0 & -1.0 & n    V
15 & BD $-$14 5040     &  $-$6.93       &     0.80        & T   & G08,P15,M16      \\ % & 15.0 &  1.0 & 15.0 & -1.0 & (n)  V
16 & $\lambda$ Cep     &  $-$3.99       &  $-$7.08        & H   & B61,H01,G11,v88  \\ % & -2.0 & -3.0 &  0.0 & -3.0 & (n)  I
17 & HD 152\,623 AaAbB &  $-$6.28       &     1.44        & H   & S96,M05b,M09,M16 \\ % &  2.0 &  2.0 &  0.0 & -3.0 & (n)  V
18 & HD 201\,345       &  $-$6.08       &     1.57        & H   & C74,d05,S08,T11  \\ % & 15.0 & -6.0 & -5.0 & -1.0 &             IV
19 & $\alpha$ Cam      &     4.87       &  $-$5.23        & H   & B61,d85,S91,M05b \\ % & 13.0 & -4.0 & -3.0 &  1.0 &             Ia
20 & HD 75\,222        &     3.91       &  $-$5.92        & T   & H01,S08,T11,S14  \\ % & -5.0 & -4.0 &  5.0 &  1.0 &             Iab
21 & HD 175\,876       &  $-$5.80       &  $-$0.67        & H   & S96,S08,T11,S14  \\ % &-15.0 &  1.0 &  5.0 &  1.0 & (n)  III
22 & HD 153\,919       &     2.03       &     7.47        & T   & S75,A01,M04,S08  \\ % & -5.0 &  1.0 & -3.0 &  2.0 &             Ia
23 & HD 192\,281       &     5.72       &     0.34        & T   & C74,M04,S08,T11  \\ % & -4.0 &  1.0 & -5.0 &  1.0 & (n)  IV
24 & BD $-$08 4617     &  $-$0.10       &     7.76        & T   & M04              \\ % & -5.0 &  1.0 & 17.0 &  1.0 & (n)  III
25 & HD 14\,633 AaAb   &  $-$4.91       &     3.24        & H   & S78,B05,S08,T11  \\ % & -2.0 &  2.0 &  8.0 & -3.0 &             V
26 & HD 195\,592       &     3.42       &     5.77        & T   & C74,v95,S08,T11  \\ % &  3.0 &  2.0 & 13.0 &  0.0 &             Ia
27 & 9 Sge             &  $-$4.12       &  $-$3.46        & T   & S82,N97,M04,S08  \\ % & -5.0 & -5.0 & -5.0 & -1.0 &             Iab
28 & HD 96\,917        &     4.61       &     0.09        & T   & T11,S14          \\ % &  5.0 &  4.0 &  5.0 &  1.0 & (n)  Ib
29 & $\xi$ Per         &     4.54       &     0.14        & H   & B61,H01,T11,S11  \\ % & -5.0 &  1.0 & -3.0 &  1.0 & (n)  III
\hline
\end{tabular}
\end{center}
A01:  \citet{Ankaetal01},  B54:  \citet{BlaaMorg54},  B61:  \citet{Blaa61},      B01:  \citet{Bagnetal01},  
B05:  \citet{Boyaetal05},  C74:  \citet{CruGetal74},  d85:  \citet{deVr85},      d05:  \citet{deWietal05},  
D12:  \citet{deBrEile12},  G79:  \citet{GullSofi79},  G08:  \citet{GvarBoma08b}, G11:  \citet{GvarGual11},
G12:  \citet{Grunetal12},  G85:  \citet{GiesBolt85},  H01:  \citet{Hoogetal01},  L12:  \citet{LopSetal12}, 
M98:  \citet{Moffetal98},  M04:  \citet{Mdzi04},      M05a: \citet{MdziChar05},  M05b: \citet{Meuretal05},  
M09:  \citet{Masoetal09},  M12:  \citet{Moldetal12},  M16:  \citet{Maizetal16},  N97:  \citet{NorCetal97},  
P15:  \citet{Perietal15},  R02:  \citet{Riboetal02},  S75:  \citet{Suta75},      S78:  \citet{Ston78},      
S82:  \citet{Ston82},      S91:  \citet{Ston91},      S96:  \citet{Sayeetal96},  S08:  \citet{SchiRose08},  
S11:  \citet{Sotaetal11a}, S14:  \citet{Sotaetal14},  T10:  \citet{Tetzetal10},  T11:  \citet{Tetzetal11},  
v88:  \citet{vanBMcCr88},  v95:  \citet{vanBetal95},  W89:  \citet{WisoWend89}.
\end{table*}

$\,\!$\indent The O stars that were known to be runaways before Paper I was published are given in Table~\ref{tabs1}. Many
of the stars are very bright, as evidenced by the fact that almost half of this sample has no TGAS proper motions. A list of 
references is provided for each object. The list is not intended to be complete, as that would occupy too much space (the 
list includes some of the first objects identified as runaway stars), but includes the first reference to the runaway 
character of each object that we have found. We will discuss this sample later on when dealing with completeness.

All of the objects in Table~\ref{tabs1} were presented in GOSSS I+II+III (the reader is referred to those papers for a brief
description of each star) except for BD~$-$08~4617 (= ALS~9668 = LS~IV~$-$08~7), which is added to
the GOSSS main catalog here. It was classified as O8.5~V: by \citet{Morgetal55}. Our spectrogram in 
Fig.~\ref{GOSSS_O} coincides with their spectral subtype but assigns a giant luminosity class and an (n) rotation index.

\subsection{New runaways from Paper I}

\begin{table}
\caption{New O- and B0-type runaway star candidates from Table~1 in Paper~1. The list is sorted
by $\Delta$, the normalized deviation from the mean latitude and longitude proper motions for their 
Galactic longitude. An ID is provided for each star in order to identify it in Figs.~\ref{pmlat}~and~\ref{pmlon}. The corrected 
proper motions are in mas/a. The T/H flag indicates the origin of the proper motions (TGAS or Hipparcos, 
respectively). 
See Tables~\ref{tabs1},~\ref{tabs3a},~and~\ref{tabs3b} for the rest of the runaways in this paper.} % CHANGE
\label{tabs2}
\begin{center}
\begin{tabular}{clrrc}
\hline
ID & Name            & $\mu_b^\prime$ & $\mu_l^\prime$  & T/H \\ 
\hline
A  & HD 155\,913     &   12.08        &    0.35         & T   \\  % & -7.0 &  2.0 & -5.0 &  1.0 & NGC 6322                \\ n    V
B  & HD 104\,565     &    4.10        & $-$8.66         & T   \\  % &  0.0 & -6.0 &  5.0 &  1.0 & GP, $4.0^{\rm o}$ away  \\           Iab
C  & ALS 18\,929$^*$ &    7.08        & $-$0.99         & T   \\  % & -3.0 &  1.0 & -5.0 &  1.0 & GP, $10.6^{\rm o}$ away \\ taken away from sample
D  & ALS 11\,244     &    6.33        & $-$3.24         & T   \\  % &  5.0 & -5.0 &  5.0 &  1.0 & Cyg OB2                 \\ (n)  III
E  & HDE 229\,232 AB &    4.44        &    6.49         & T   \\  % & -5.0 &  2.0 & -3.0 &  0.0 & NGC 6913                \\ n    V
F  & HD 155\,775     & $-$6.26        & $-$0.93         & T   \\  % & 10.0 &  5.0 &  6.0 & -3.0 & ---                     \\ (n)  III
G  & HD 46\,573      & $-$5.03        & $-$4.77         & T   \\  % & -5.0 & -4.0 &  6.0 & -3.0 & GP, $2.6^{\rm o}$ away  \\           V
H  & BD +60 134      & $-$5.44        &    2.01         & T   \\  % & -5.0 &  1.0 & -5.0 & -3.0 & Cas OB7                 \\ (n)  V
I  & CPD -34 2135    &    3.95        &    4.76         & T   \\  % & -5.0 &  1.0 & -5.0 &  1.0 & ---                     \\           Ib
J  & HD 12\,323      & $-$5.02        &    2.07         & T   \\  % & -5.0 &  1.0 &  0.0 & -3.0 & Per OB1                 \\           V
K  & AB Cru          &    0.36        & $-$6.69         & T   \\  % & 10.0 & -1.0 & -4.0 &  2.0 & ---                     \\           III
L  & HD 192\,639     &    3.84        &    3.39         & T   \\  % & -1.0 &  2.0 & -3.0 & -2.0 & Dolidze 4               \\           Iab
M  & HD 94\,024      &    4.46        & $-$0.52         & T   \\  % & -3.0 & -7.0 & -5.0 &  1.0 & Carina Nebula           \\           IV
\hline
\end{tabular}
\end{center}
$^*$ Not an O star but a B0~V.
\end{table}

$\,\!$\indent The stars identified as runaways for the first time in Paper I are given in Table~\ref{tabs1}. Each one is briefly described in this
subsection, ordered by decreasing $\Delta$ or likelihood of being a runaway from their corrected proper motions.

\paragraph{HD 155\,913 (= CPD~$-$42~7710 = ALS~4407).} GOSSS II classified this star as O4.5~Vn((f)) but cautioned that it is an SB2
according to OWN data \citep{Barbetal10,Barbetal17}, so the width could be caused by an unresolved orbital motion in the GOSSS spectrogram. 
\citet{Aldoetal15} also detected a visual companion.
The proper motion of the star points away from the young stellar cluster NGC~6822, located half a degree away and the likely source of this 
runaway star. No bow shock is visible in the {\it WISE} images but the region of the sky is quite crowded due to its location.

\paragraph{HD 104\,565 (= CPD~$-$57~5199 = ALS~2572).} This nitrogen-deficient star was classified as OC9.7~Iab in GOSSS-II.
HD 104\,565 is 4\degr\ above the Galactic Plane and moving away from it and towards the East. Its proper motion points away from Cen~OB1, 
$\sim$8\degr\ away and a possible origin for this runaway. 
%No warm dust is visible in its neighborhood in the WISE images.

\paragraph{ALS 18\,929 (= LSE~107 = Tyc~1036-00450-1).} This object was given an O9.7 spectral subtype in GOSSS III with no luminosity class 
because of the discrepancy between the He/He and Si/He criteria. With the help of new data we have reanalyzed the spectral
classification and determined that B0~V is more appropriate, as \SiIII{4552} is stronger than \HeII{4542}, see Fig.~\ref{GOSSS_B}. Therefore, 
we move this object from the main GOSC catalog (Galactic O stars) to supplement 2 (other Galactic early-type stars). It is still likely to be a
runaway, as it is a massive star located 10\fdg6 off from the Galactic Plane and moving away from it in a direction close to its perpendicular. 
%No warm dust is visible in its neighborhood in the WISE images.

\paragraph{ALS 11\,244 (= LS~III~+41~20).} \citet{ComePasq12} classified this star as O5~If. The spectrogram in Fig.~\ref{GOSSS_O} indicates 
a similar spectral subtype (O4.5) but a lower luminosity class of III, as \HeII{4686} is not clearly in emission. Instead, it shows the double
emission peak surrounding an absorption line characteristic of Onfp stars (see GOSSS-I). It also has \CIII{4647-50-52} comparable to 
\NIII{4634-40-42}, hence the (fc) suffix in the classification \citep{Walbetal10a}.
Based on its proper motion, the most likely origin for this runaway star is the OB association Cyg~OB2, 
located $\sim$2\degr\ away. 
%The region of the sky where ALS~11\,244 currently is includes the MIR-bright wall of a cavity, so it is not easy to
%discern a possible bow shock in the WISE images.

\paragraph{HDE 229\,232 AB (= BD~$+$38~4070 = ALS~11\,296 = LS~II~$+$38~79).} GOSSS-III classified this star as O4~V:n((f)). The object has a
bright visual companion \citep{Aldoetal15} and is an SB1 \citep{Willetal13}. Based on its proper motion, the most likely origin of this 
runaway star is the young cluster NGC~6913, located $\sim$40\arcmin\ away, but note that HDE~229\,232~AB is much earlier (hence, likely more massive)
than the cluster turnoff \citep{Negu04}. Some warm dust is visible in the {\it WISE} images around HDE 229\,232 AB
(Fig.~\ref{WISE}) but it does not appear to originate in a bow shock.

\paragraph{HD 155\,775 (= V1012~Sco = CPD~$-$38~6750 = ALS~3995).} GOSSS-II classified this object as O9.7~III(n) and \citet{Malketal06} 
indicates it is an eclipsing binary. It is located close to the Galactic Plane and moving away from the open cluster ASCC 88, which could be its
origin. The {\it WISE} image (Fig.~\ref{WISE}) shows a weak bow shock consistent with the direction of the corrected proper
motion but note the region is quite crowded \citep{Kobuetal16}. The Herschel~160~$\mu$m band also shows a small cavity around the star consistent 
with the bow shock structure.
%Status: Previously known O runaway (since the IAU contribution)
%Bow shock: We have possibly found a bow shock in the direction of motion of the system. It can be seen on high contrast W4 band images from WISE. 
%It could also be a photodissociation region. In Herschel's 160 micron band we see a small structure (a cavity) surrounding the star, occupying the 
%same volume as the candidate bow shock.
%Discussion: E-mail thread called "¿Onda de proa?" between Jesus Maíz, Rodolfo Barbá and Michelangelo Pantaleoni

\paragraph{HD 46\,573 (= BD~$+$02~1295 = ALS~9029 = LS~IV~$+$02~10).} GOSSS-I classified this star as O7~V((f))z. The object is moving away from the
Galactic Plane and tracing back its trajectory we arrive at NGC~6822, an \HII\ region with a young cluster and a potential origin for the runaway.
Note that another runaway, HD~47\,432, is present nearby. We do not detect it with our method (i.e. it is one of our false negatives) 
but it is identified as a runaway by \citet{Tetzetal11}. The
trajectories of both runaways intersect at a near right angle at the location of Collinder 110, but that open cluster is too old to have produced
such young stars, so it is likely a chance superposition (we will check this with Gaia DR2, nonetheless). The {\it WISE} image shows a strong bow shock 
consistent with the corrected TGAS proper motion (Fig.~\ref{WISE}, see Paper I and \citealt{Kobuetal16}). 

\paragraph{BD~$+$60~134 (= ALS~6405 = LS~I~$+$61~173).} GOSSS III obtained a spectral classification of O5.5 V(n)((f)) for this star.
BD~$+$60~134 is in the neighborhood of several OB associations in Cassiopeia. Based on its 
proper motion, Cas~OB7 appears to be the most likely origin, but note that the star is earlier (hence, likely more massive)
than any star in the association.
%No structure is observed around it in the W4 WISE image.

\paragraph{CPD $-$34 2135 (= ALS~1007).} \citet{Garretal77} classified this star as O7~III. We find a slightly later spectral subtype and a brighter 
luminosity class, O7.5 Ib(f)p. The p suffix is assigned for its incipient P-Cygni profile in \HeII{4686}, see Fig.~\ref{GOSSS_O}.
CPD~$-$34~2135 is located very close to the Galactic Plane and moving in a diagonal direction with respect to it. Tracing back its proper motion,
a possible origin in the young star clusters in Canis Majoris is found. 
%No warm dust is visible in its neighborhood in the WISE images.
% (change from O7.5 II-Ib(f)p).

\paragraph{HD 12\,323 (= BD~$+$54~441 = ALS~6886 = LS~I~$+$55~22).} This nitrogen-rich star was classified as ON9.2~V in GOSSS II. \citet{MusaChen89}
found it to be a spectroscopic binary and \citet{Kendetal95} identified it as a blue straggler. This object is over 5 degrees to the south of the 
Galactic Plane and moving away from it. Its proper motion points towards the OB associations Per~OB1 and Cas~OB8 as possible origins. 
%No warm dust is visible in its neighborhood in the WISE images.

\paragraph{AB Cru (= HD~106\,871 = CPD~$-$57~4397 = ALS~2639).} \citet{Garretal77} classified this star as O8~Vn but other sources identify it 
as a B star. One possible reason for the discrepancies is that the spectrum of this object is highly variable due to its eclipsing binarity
\citep{Popp66}. Here we use GOSSS spectra to determine it is an O8.5 III + BN0.2: Ib: system (Fig.~\ref{GOSSS_O}), one of the very few known 
combinations of an O star and a B supergiant in a short-period binary\footnote{But note that the luminosity class of the B star is uncertain.}. 
The B supergiant has a significant N enrichment. We are using GOSSS and 
OWN \citep{Barbetal10,Barbetal17} spectroscopy to determine the orbit. AB Cru is 4\fdg4 above the Galactic Plane and moving westward almost parallel to it. 
If it has been ejected from a system to the East, then its motion is apparently being already bent back towards the Galactic Plane\footnote{Throughout 
this paper we will say that a runaway star is moving towards the Galactic Plane when it is doing it on the plane of the sky i.e. when 
$\mu_b^\prime$ and $b$ have opposite signs. As we do not have information on the radial velocity and the distance, we do not know if that is also true 
in terms of true 3-D velocity.}. One candidate for its origin is the Cen~OB1 association, $\sim$6\degr\ away. 
%No warm dust is visible in its neighborhood in the WISE images.
% (change from O8 III + B0.7: Ib:).

\paragraph{HD 192\,639 (= BD~$+$36~3958 = ALS~10\,996 = LS~II~$+$37~26).} GOSSS-I classified this star as O7.5~Iabf (actually, it is an O7.5~Iab
standard). It is located within the Cyg~OB1 association moving northwards away from the Galactic Plane. In paper I we proposed the open cluster
Dolidze~4 as a possible origin but that is uncertain, as the corrected proper motion traces back to a point close to the cluster but not within it.

\paragraph{HD 94\,024 (= CPD~$-$57~3856 = ALS~1952).} GOSSS-II classified this star as O8~IV. OWN data \citep{Barbetal10,Barbetal17} indicates this 
system is an SB1. HD 94\,024 is moving away from the Carina Nebula Association, its likely origin, located 2\degr\ to the south.

\subsection{Objects not in Paper I}

$\,\!$\indent In this subsection we describe the runaway candidates that were not discussed in Paper I, dividing them in O stars
%, WRs, 
and BA supergiants.

\subsubsection{O stars}

\begin{table}
\caption{O-type runaway star candidates not in Paper~1. The list is sorted
by $\Delta$, the normalized deviation from the mean latitude and longitude proper motions for their 
Galactic longitude. An ID is provided for each star in order to identify it in Figs.~\ref{pmlat}~and~\ref{pmlon}.
The corrected proper motions are in mas/a. The T/H flag 
indicates the origin of the proper motions (TGAS or Hipparcos, respectively). The P/S flag 
indicates whether the identification of the star as being of O type is photometric or
spectroscopic, respectively. The Known column indicates whether the runaway character of the
object had been identified previously or not. 
See Tables~\ref{tabs1},~\ref{tabs2},~and~\ref{tabs3b} for the rest of the runaways in this paper.} % CHANGE
\label{tabs3a}
\begin{center}
\begin{tabular}{clrrccc}
\hline
ID            & Name             & $\mu_b^\prime$ & $\mu_l^\prime$  & T/H & P/S & Known \\
\hline
$\alpha$      & HDE 315\,927     &    5.27        & $-$6.46         & T   & P   & no    \\ % & -1.0 & -4.0 & -5.0 &  1.0 % No GOSSS
$\beta$       & CPD $-$72 1184   & $-$7.04        &    0.37         & H   & P   & yes   \\ % &  2.0 &  4.0 & -5.0 & -2.0 % No GOSSS
$\gamma$      & ALS 4962         & $-$5.47        &    5.66         & T   & S   & no    \\ % & -5.0 &  1.0 & -5.0 & -1.0 %                  I
$\delta$      & CPD $-$63 2886   &    0.63        &    9.02         & T   & P   & no    \\ % & 10.0 & -1.0 &  1.0 &  3.0 % No GOSSS
%$\varepsilon$ & Tyc 3159-00006-1 &    2.58        &    7.53         & T   & S   & yes   \\
$\varepsilon$ & HDE 322\,987     & $-$4.29        & $-$3.73         & T   & P   & no    \\ % & -1.0 & -4.0 & -5.0 & -1.0 % No GOSSS
$\zeta$       & CPD $-$57 3781   &    4.58        & $-$0.11         & T   & P   & yes   \\ % &  4.0 & -8.0 & -5.0 & -1.5 % No GOSSS
$\eta$        & HD 74\,920       &    0.17        & $-$6.12         & T   & S   & no    \\ % &  5.0 & -4.0 &  2.0 &  3.0 % n          IV
\hline
\end{tabular}
\end{center}
\end{table}

$\,\!$\indent The (spectroscopically confirmed or not) O stars in this subsection are listed in Table~\ref{tabs3a}. Each one is briefly described in this
subsubsection, ordered by decreasing $\Delta$.

\paragraph{HDE 315\,927 (= CPD~$-$29~4760 = ALS~4200).} \citet{VijaDril93} classified this star as O5~III but we do not have GOSSS data to verify it.
The CHORIZOS run (Table~\ref{chorizos}) yields a \teff\ consistent with that spectral classification and a moderate extinction with an \rv\ close to
the canonical value of 3.1. We have not found any previous identifications of HDE~315\,927 as a runaway star. Tracing back its proper motion leads 
to several clusters in Sgr in the direction of the Galactic Center as its possible origin. 

\paragraph{CPD~$-$72~1184 (= ALS~2557).} This object has been classified as a late-type O (O9~III, \citealt{Wram80}) or as an early-type B 
(B0~III, \citealt{Kilk74}). We do not have GOSSS data to verify the spectral classification and the CHORIZOS run (Table~\ref{chorizos}) yields a 
\teff\ consistent with a late-type O but without excluding the possibility of being an early-type B and a low extinction. Therefore, we tentatively 
leave it here with the intention of confirming the spectral classification in the near future. \citet{Kilk74} identified it as a runaway star and 
\citet{deBrEile12} measured a radial velocity of $-$218~km/s. CPD~$-$72~1184 is located 11\degr\ to the south of the Galactic Plane and moving in a 
direction almost perpendicular to it. It could have originated in the region between Centaurus and Crux. This is the only object in this subsubsection 
with a Hipparcos instead of a TGAS proper motion.

\paragraph{ALS~4962.} This star is potentially the most interesting object in this subsubsection. It has received little previous attention but
\citet{Kilk93} classified it as O9~Ia. The GOSSS spectrum confirms its supergiant nature (but note that \HeII{4686} is not fully in emission and is
closer to a P-Cygni profile) and brings the spectral subtype to a much earlier O5. O5 is established, as usual, from the \HeII{4542}/\HeI{4471} 
ratio but that leaves us with N\,{\sc iii}+{\sc iv}+{\sc v} absorption lines significantly stronger than expected, hence the ON designation and the 
p suffix, as the ON phenomenon is not expected to appear in early/mid supergiants. It is also possible that the spectrum is a composite of an earlier
O and a later O stars but even under that assumption it would be hard to reach the intensity of the nitrogen absorption lines. ALS~4962 shows strong
atomic, molecular, and DIB ISM absorption lines, an indication of strong extinction. We have found no previous indication of its runaway nature in the
literature. There is a faint structure in the {\it WISE} W4 band consistent with a bow shock slightly offset from the direction of the corrected proper 
motion (Fig.~\ref{WISE}). Tracing back its motion leads to the western part of the Sgr~OB1 association near the location of M20 as a possible origin.

\paragraph{CPD~$-$63~2886 (= ALS~3140).} \citet{VijaDril93} classified this star as O9.5~Ib but we do not have GOSSS data to verify it.
The CHORIZOS run (Table~\ref{chorizos}) yields a \teff\ consistent with that of an O star, likely even earlier than O9.5, and a relatively high
extinction with an \rv\ close to the canonical value of 3.1. We have not found any previous identifications of CPD~$-$63~2886 as a runaway star. 
It is moving eastwards in a direction parallel to the Galactic Plane away from the Cen~OB1 association, a possible origin. 

%\paragraph{Tyc 3159-00006-1.} \citet{Gvaretal14} classified this object as O9.5-9.7~Ib and identified it as a runaway star. The GOSSS spectral type is
%only slightly different, O9.7~Iab.

\paragraph{HDE 322\,987 (= CPD~$-$37~7076 = ALS~4059).} This object appears as O7 in \citet{Goy73} and as O5~V in \citet{VijaDril93} but we do not 
have GOSSS data to verify the spectral classification. The CHORIZOS run (Table~\ref{chorizos}) yields a \teff\ consistent with those classifications 
along with a moderate extinction with an \rv\ close to the canonical value of 3.1. We have found no previous indication of its runaway nature in the
literature. The {\it WISE} mosaic shows extended emission in the region expected for a bow shock (Fig.~\ref{WISE} and \citealt{Kobuetal16}) but the 
region is crowded and other sources are present, so an alternative explanation such as a PDR cannot be discarded. Regarding possible origins, this is 
a crowded region where it is difficult to select one without further information. One of the candidates is the \HII\ region RCW~126.
%Status: Possible unknown O runaway
%Bow shock: A structure resembling a bow shock appears in WISE images taken with the W4 filter. The orientation of the bow shock candidate is in 
%good agreement with what should be expected from the proper motion vector from TGAS. Since the star lies in the direction of the galactic center 
%the infrared emission in the WISE images should be regarded with caution since a PDR in the high density interstellar medium could still be a 
%possibility.
%Discussion: E-mail thread called "¿Otra onda de proa?" between Jesus Maíz, Rodolfo Barbá and Michelangelo Pantaleoni

\paragraph{CPD~$-$57~3781 (= ALS~1887).} \citet{Cram71} classified this star as O8 and \citet{VijaDril93} as O7~V(n) but we do not 
have GOSSS data to verify the spectral classification. The CHORIZOS run (Table~\ref{chorizos}) yields a \teff\ consistent with those classifications 
and a moderate extinction with a value of \rv\ significantly higher than the canonical one. 
This object is moving away form the Carina Nebula association, located 1\degr\ away, its more likely origin. CPD~$-$57~3781 is the 
ionizing star of the optical \HII\ region RCW~52, whose shape and alignment is consistent with being at least partially a bow shock 
created by this runaway star, a fact first noted by \citet{Cappetal05}.

\paragraph{HD 74\,920 (= CPD~$-$45~2978 = ALS~1148).} GOSSS-II classified this star as O7.5~IVn. It was not included in paper I as it was just below
the threshold there but with the slightly modified parameters here it is just above it. We have not found any previous identifications of this object
as a runaway star. HD~74\,920 is moving westwards in a direction nearly parallel to the Galactic Plane within the Vel~OB1 association. It could have
originated in one of the \HII\ regions towards the east such as RCW~38 or RCW~40.

%\subsubsection{WR stars}
%
%$\,\!$\indent Our analysis for Wolf-Rayet stars has revealed only three runaways, all of them previously known. They are WR~2 \citep{Tetzetal11},
%WR~71 \citep{Moffetal98}, and WR~93 \citep{Gvaretal11b}.

\subsubsection{BA supergiants}

\begin{table}
\caption{BA-supergiant runaway star candidates. The list is sorted
by $\Delta$, the normalized deviation from the mean latitude and longitude proper motions for their 
Galactic longitude. An ID is provided for each star in order to identify it in Figs.~\ref{pmlat}~and~\ref{pmlon}.
The corrected proper motions are in mas/a. The T/H flag 
indicates the origin of the proper motions (TGAS or Hipparcos, respectively). 
The Known column indicates whether the runaway character of the
object had been identified previously or not. The last column indicates whether the spectral type is
E(arly, from B0 to B0.7) or L(ate, B1 or later). 
See Tables~\ref{tabs1},~\ref{tabs2},~and~\ref{tabs3a} for the rest of the runaways in this paper.} % CHANGE
\label{tabs3b}
\begin{center}
\begin{tabular}{clrrccc}
\hline
ID & Name             & $\mu_b^\prime$ & $\mu_l^\prime$  & T/H & Known   & E/L \\
\hline
a  & HD 150\,898      &  $-$8.46       & $-$15.37        & H   & yes     & E   \\ % & -5.0 & -1.0 & -3.0 & -3.0
b  & 69 Cyg           & $-$10.61       &     4.40        & T   & yes     & E   \\ % &  3.0 &  2.0 & -4.0 & -2.0
c  & HD 215\,733      &  $-$9.56       &  $-$6.61        & H   & yes     & L   \\ % &  7.0 & -3.0 & -4.0 &  0.0
d  & $\gamma$ Ara     &  $-$7.18       & $-$10.62        & H   & no      & L   \\ % &  8.0 & -3.0 &  0.0 &  1.0
e  & HD 119\,608      &     8.17       &  $-$6.49        & H   & unclear & L   \\ % &  7.0 & -3.0 & -3.0 &  0.0
f  & HD 156\,359      &  $-$8.53       &  $-$3.51        & H   & yes     & E   \\ % &  3.0 & -5.0 & -5.0 &  0.0
g  & BD $-$08 4623    &     8.09       &  $-$2.03        & T   & no      & E   \\ % &  4.0 & -5.0 & -5.0 &  1.0
h  & HD 86\,606       &  $-$5.67       &  $-$8.07        & T   & yes     & L   \\ % &  4.0 & -5.0 &  4.0 & -3.0
i  & $\theta$ Ara     &     4.51       &  $-$9.11        & H   & yes     & L   \\ % &  7.0 & -3.0 & -5.0 &  1.0
j  & HD 112\,272      &  $-$6.88       &  $-$3.36        & T   & yes     & E   \\ % &  4.0 & -5.0 &  4.0 & -3.0
k  & $\rho$ Leo AB    &  $-$5.94       &     2.00        & H   & yes     & L   \\ % &  4.0 &  3.0 & -5.0 & -1.0
l  & $\phi$ Vel       &  $-$4.35       &  $-$6.20        & H   & no      & L   \\ % & -5.0 & -1.0 & -5.0 & -1.0
m  & HD 125\,288      &  $-$2.89       &  $-$7.10        & H   & no      & L   \\ % &  4.0 & -5.0 & -3.0 & -2.0
n  & 67 Oph           &  $-$4.41       &  $-$4.59        & H   & no      & L   \\ % &  3.0 & -5.0 &  7.0 & -3.0
o  & HD 167\,756      &  $-$5.27       &  $-$0.80        & H   & no      & E   \\ % &  3.0 &  5.0 &  5.0 & -3.0
p  & HD 161\,695      &  $-$4.06       &     4.87        & T   & no      & L   \\ % & -5.0 &  1.0 &  4.0 & -3.0
q  & HD 115\,842      &     4.80       &     1.92        & T   & no      & E   \\ % & -3.0 &  2.0 & -5.0 &  1.0
r  & CPD $-$50 9557   &     0.64       &  $-$7.17        & T   & no      & E   \\ % &  6.0 & -1.0 & -1.0 &  2.0
s  & $\kappa$ Cas     &  $-$1.45       &     6.52        & H   & yes     & E   \\ % & -5.0 &  1.0 &  3.0 & -4.0
t  & HD 8065          &     4.16       &  $-$2.62        & T   & unclear & L   \\ % &  4.0 & -5.0 & -5.0 &  1.0
u  & GP Vel           &     3.34       &  $-$4.52        & T   & yes     & E   \\ % & -5.0 & -2.0 & -5.0 &  1.0
v  & HD 190\,066      &  $-$1.09       &  $-$6.54        & T   & yes     & E   \\ % & -5.0 & -1.0 & -6.0 &  2.5
w  & V452 Sct         &     2.01       &  $-$5.10        & T   & yes     & L   \\ % &  0.0 & -4.0 & -5.0 &  1.5
x  & HD 94\,909       &     3.90       &  $-$0.74        & T   & no      & E   \\ % & -4.0 &  2.0 & -5.0 &  1.0
y  & HD 171\,012      &  $-$2.01       &  $-$5.01        & H   & yes     & E   \\ % & -4.0 & -4.0 &  3.0 & -3.0
z  & HD 161\,961      &     3.58       &     2.23        & T   & no      & E   \\ % &  5.0 &  2.0 &  4.0 &  1.0
aa & HD 190\,603      &  $-$1.93       &  $-$5.04        & T   & no      & L   \\ % &  7.0 & -4.0 &  7.0 & -4.0
\hline
\end{tabular}
\end{center}
\end{table}

$\,\!$\indent The BA supergiants in this subsection are listed in Table~\ref{tabs3b}. Each one is briefly described in this
subsubsection, ordered by decreasing $\Delta$.

\paragraph{HD~150\,898 (= CPD~$-$58~693 = ALS~15\,035).} GOSSS data yields a spectral type of B0~Ib for this star, which had been previously 
identified as a runaway \citep{Hoogetal01,Tetzetal11}. A clear bow shock is visible in the {\it WISE} mosaic (Fig.~\ref{WISE}). Its proper motion traces 
back to the Ara-Scorpius region of the Galactic Plane 15-20\degr\ away.

\paragraph{69~Cyg (= HD~204\,172 = BD~$+$36~4557 = ALS~14\,838 = V2157~Cyg).} This object has a spectral type of B0.2~Ib in GOSSS data and had been 
previously identified as a runaway \citep{Hoogetal01,MdziChar05}. Tracing back its proper motion yields a likely origin in the Cygnus region
$\sim$10\degr\ away.

\paragraph{HD~215\,733 (= BD~$+$16~4814 = ALS~14\,757).} GOSSS data yields a spectral type of B1~Ib for this star, which had been previously 
identified as a runaway \citep{Tetzetal11}. Based on its proper motion, a possible origin is in the Lac~OB1 or Cep~OB2 associations 30-40\degr\ away.

\paragraph{$\gamma$~Ara (= HD~157\,246 = CPD~$-$56~8225 = ALS~15\,048).} This object has a spectral type of B1~Ib in GOSSS data and we have not found 
any reference to its runaway character. It is one of the closest objects in our list but both its Hipparcos and spectroscopic parallaxes place it
beyond 300~pc. Its proper motion traces back to the Scorpius region of the Galactic Plane 20-25\degr\ away.

\paragraph{HD~119\,608 (= BD~$-$17~3918 = ALS~14\,746).} \citet{Morgetal55} classified this star as B1~Ib but we have not observed it with GOSSS.
It is not clear whether this object had been previously identified as a runaway: \citet{McEvetal17} indicates that it is one but use \citet{Mart04}
as a reference for that claim. What  
\citet{Mart04} % CHANGE
actually says is that HD~119\,608 is a 
Post Asymptotic Giant Branch (PAGB) star % CHANGE
instead of a runaway B supergiant, while the latter 
\cite{Szczetal07} says it is not a PAGB, something that unpublished data from the IACOB project \citep{SimDetal15b} also indicates. 
HD~119\,608 is located far above the Galactic Plane and its possible origin could have been the Sgr region of the Galactic Plane,
$\sim$60\degr\ away.

\paragraph{HD~156\,359 (= CPD~$-$62~5531 = ALS~16\,942).} GOSSS data yields a spectral type of B0~Ia for this star, which had been previously 
identified as a runaway \citep{HousKilk78,Mdzi04}. Tracing back its corrected proper motion leads back to Ara~OB1 as a possible origin.

\paragraph{BD~$-$08~4623 (= ALS~9685 = LS~IV~$-$08~8).} GOSSS yields a spectral type of B0.5: Ia:, with the uncertainty in the classification 
arising from the relative low S/N of the data (see~Fig.\ref{GOSSS_B}). There are also hints that the spectrum could be composite: we will observe
it again in the future to obtain better data. We have found no indication of its runaway nature in the literature. The {\it WISE} mosaic (Fig.~\ref{WISE})
shows an asymmetrical warm-dust region around the star as opposed to a clear bow shock (although the position is the expected one from the corrected
proper motion) and the star has apparently created a bubble around it \citep{Simpetal12}. 
BD~$-$08~4623 is in the Sct region of the Galactic Plane moving towards the north, so it is likely that its origin is nearby.

\paragraph{HD~86\,606 (= CPD~$-$70~953 = ALS~14\,941).} This object has a spectral type of B1~Ib in GOSSS data and was identified as a runaway star
by \citet{Tetzetal11}. According to the corrected proper motion a possible origin is Cen~OB1, $\sim$18\degr\ away.

\paragraph{$\theta$~Ara (= HD~165\,024 = CPD~$-$50~10538 = ALS~15\,053).} \citet{Hiltetal69} classifies it as B2~Ib but Simbad lists other
classifications ranging from B0.5~Iab to B1~II. We have not obtained GOSSS data for this star, which is one of the closest targets in our sample. 
It was identified as a runaway star by \citet{Tetzetal11}. It is already falling back towards the Galactic Plane, so its origin is hard to trace based
only on the corrected proper motion.

\paragraph{HD~112\,272 (= DW~Cru = CPD~$-$63~2454 = ALS~2834).} \citet{Morgetal55} classified this star as B0.5~Ia but we have not obtained a
GOSSS spectrum to confirm it. It was identified as a runaway star by \citet{Tetzetal11}. It is located within the Cen~OB1 association, moving away 
from its center.

\paragraph{$\rho$~Leo~AB (= HD~91\,316~AB = BD~$+$10~2166~AB = ALS~14\,811~AB).} GOSSS yields a spectral classification of B1~IbNstr for this
nitrogen-enhanced star, which was considered a runaway star already by \citet{KeenDuft83}. $\rho$~Leo~AB is located far away from the Galactic
Plane and already falling back towards it. The AB designation is due to the existence of an elusive bright companion (there are several historical
references that failed to detect it) that according to \citet{Tokoetal10} is 1.5 magnitudes dimmer than the primary (hence contributing significantly
to the spectrum) and separated by 46 mas. There are also claims in the literature that the object is a spectroscopic binary but an extensive 
spectroscopic monitoring performed by the IACOB project \citep{SimDetal15b} clearly demonstrates that the spectroscopic variability is 
actually produced by stellar oscillations and rotational modulation of a non-spherically symmetric wind (see also \citealt{Aertetal18}).

\paragraph{$\phi$~Vel (= HD~86\,440 = CPD~$-$53~3075 = ALS~14\,940).} \cite{Hiltetal69} classified this star as B5~Ib but we have not obtained a
GOSSS spectrum to confirm it. There is no previous reference to its runaway nature in the literature. It is located close to the Galactic Plane and
one possible origin is the open cluster ASCC~58, 3\degr\ away.

\paragraph{HD~125\,288 (= v~Cen = CPD~$-$55~5984 = ALS~14\,996).} \cite{Hiltetal69} classified this star as B6~Ib but we have not obtained a
GOSSS spectrum to confirm it. We have not found any previous reference to its runaway nature. It is moving westwards and already falling back towards the 
Galactic Plane.

\paragraph{67~Oph (= HD~164\,353 = BD~$+$02~2186 = ALS~14\,022).} The GOSSS spectral classification for this star is B5~Ib and there is no previous
mention of its runaway nature in the literature. It is already falling back towards the Galactic Plane.

\paragraph{HD~167\,756 (= CPD~$-$42~8359 = ALS~14\,536).} We classify this star as B0.2~Ib with GOSSS data. We have found no previous reference to its
runaway nature. It is moving in a direction close to the perpendicular to the Galactic Plane, with one possible origin in NGC~6357 or its
neighborhood.

\paragraph{HD 161\,695 (= BD~$+$31~3090).} The GOSSS spectrum yields a classification of A0~Ib for this star, consistent with previous results.
HD~161\,695 is not only located far from the Galactic Plane (27\degr) but is also one of the runaways in this paper that is falling back towards
the Galactic Plane. Therefore, tracing back its origin is not possible without additional distance and velocity information. Somewhat surprisingly for
a sixth-magnitude supergiant at its location, we have not found any references to its runaway nature in the literature.

\paragraph{HD~115\,842 (= CPD~$-$55~5504 = ALS~3038).} The GOSSS spectral classification for this star is B0.5~Ia and there is no previous
mention of its runaway nature in the literature. The {\it WISE} mosaic (Fig.~\ref{WISE}) shows a large and strong bow shock consistent with the corrected 
proper motion. A possible origin for this runaway is Cen~OB1, located 7\degr\ away.

\paragraph{CPD~$-$50~9557 (= ALS~3689).} We classify this star as B0~Ib with GOSSS data and we have found no previous reference to its
runaway nature in the literature. It is moving westward below the Galactic Plane, so its origin is difficult to ascertain.

\paragraph{$\kappa$~Cas (= HD~2905 = BD~$+$61~102 = ALS~6258).} GOSSS yields a spectral classification of BC0.7~Ia for this object,
which was previously known to be a runaway \citep{vanBetal95,Tetzetal11}. The 
{\it WISE} mosaic (Fig.~\ref{WISE}) shows a large and strong bow shock consistent with the corrected proper motion. $\kappa$~Cas is moving eastward close 
to the Galactic Plane within the region of the sky shared by Cas~OB4 and Cas~OB14 OB associations. It is interesting to note that this is a carbon-enhanced 
B supergiant. There are several other runaways in this paper with CNO peculiarities and Sk~$-$67~2 in the LMC was recently discovered by \citet{Lennetal17} 
to be another case of a carbon-enhanced runaway B super/hypergiant.

\paragraph{HD 8065 (= BD~$+$77~49).} This supergiant has published spectral subtypes that range from B9 to A2 but a comparison of the GOSSS 
spectrum with the A0 supergiant standards indicates that it is an A0 Iab. We list the previously known status of this little-studied star as 
``unclear'' since the only relevant information we have found in the literature is a note in Table~1 of \citet{Bide88} that says ``has this
high-latitude B9~Iab star escaped from h~and~$\chi$ Persei?''. The answer to that question from TGAS data is that it is indeed a possibility: 
the corrected proper motion traces back to the Perseus double cluster 20\degr\ away (crossing the Galactic Plane in the meantime) but with such a
long distance one would need a careful 3-D analysis to confirm the suspicion. 

\paragraph{GP~Vel (= HD~77\,581  = Vel~X-1 = CPD~$-$40~3072 = ALS~1227).} The GOSSS spectral classification for this star is B0.5~Ia. with an 
incipient P-Cygni profile in H$\beta$ that is not strong enough to warrant a Ia+ luminosity class. GP~Vel is a high-mass X-ray binary in an
8.9 day orbit \citep{McCletal76} that was previously known to be a runaway star \citep{Kapeetal97,Tetzetal11}. The {\it WISE} W4 mosaic (Fig.~\ref{WISE}) 
shows a clear bow shock consistent with the corrected proper motion. Tracing back its trajectory leads to a possible origin in the eastern part of the
Vel~OB1 association.

\paragraph{HD~190\,066 (= BD~$+$21~4027 = ALS~10\,740 = LS~II~$+$22~20).} We classify this star as B0.7~Iab based on GOSSS data. It was previously
known to be a runaway \citep{Tetzetal11}. It is moving westward below the Galactic Plane with a possible origin in the Cygnus region.

\paragraph{V452~Sct (= BD~$-$13~5061 = ALS~5107 = LS~IV~$-$13~82).} We have not observed this object with GOSSS but the spectrum published by
\citet{Miroetal00} indicates it is an A0 Ia+. That paper also indicate that it is a likely runaway star. This is one of the more distant objects in
our sample, as it is likely beyond the Galactic Center. The {\it WISE} mosaic shows an excess in that band consistent with being a point source centered
on the star, so it is likely caused by the IR excess studied by \citet{Miroetal00} and not by a bow shock. V452~Sct is already falling back towards the
Galactic Plane. 

\paragraph{HD~94\,909 (= V526~Car = CPD~$-$56~4016 = ALS~2036).} The GOSSS spectral classification for this star is B0~Ia. To our knowledge, it had
not been previously identified as a runaway star. It is moving away from the Galactic Plane in a near perpendicular direction, with a possible origin
in the Carina Nebula Association, 2\degr\ away, or a point slightly towards the east.

\paragraph{HD~171\,012 (= V4358~Sgr = BD~$-$18~4994 = ALS~5086).} The GOSSS spectral classification for this star is B0.2~Ia and it was previously
known to be a runaway \citep{Tetzetal11}. One possible origin lies in the Sct~OB associations, 10-12\degr\ away.

\paragraph{HD~161\,961 (= BD~$-$02~4458 = ALS~9353 = LS~IV~$-$02~5).} The GOSSS spectrum yields a classification of B0.5~Ib for this star, which had
not been previously identified as a runaway. Tracing back its corrected proper motion leads to the Sct OB~associations $\sim$15\degr\ away.

\paragraph{HD~190\,603 (= V1768~Cyg = BD~$+$31~3925 = ALS~1079 = LS~II~$+$32~15).}  We have not observed this object with GOSSS but \citet{Lennetal92}
gives a spectral classification of B1.5~Ia+ and the star was analyzed in detail by \citet{Claretal12}. 
We have found no prior references to its runaway character. The {\it WISE} mosaic (Fig.~\ref{WISE}) shows an
asymmetric nebula around it with a direction consistent with the corrected proper motion. The nebula does not have the typical shape of a bow shock but
that could be due to the extremely high mass-loss rate of this object \citep{Lennetal92,Walbetal15a}, which could produce a more filled structure.
HD~190\,603 is moving westwards close to the Galactic Plane and its likely origin is in one of the Cygnus OB associations.

\section{Analysis}

\begin{figure*}
\centerline{\includegraphics[width=0.49\linewidth]{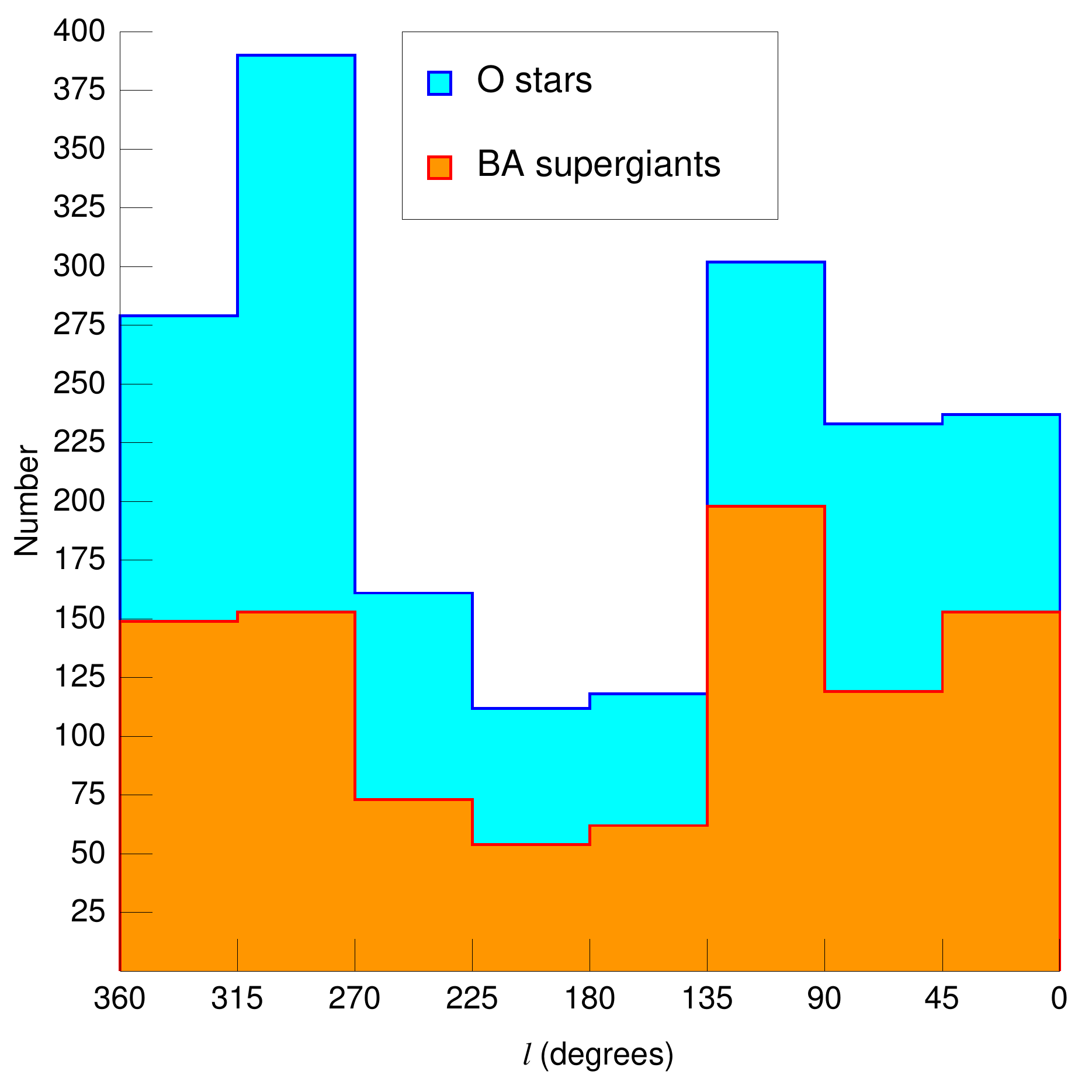} \
            \includegraphics[width=0.49\linewidth]{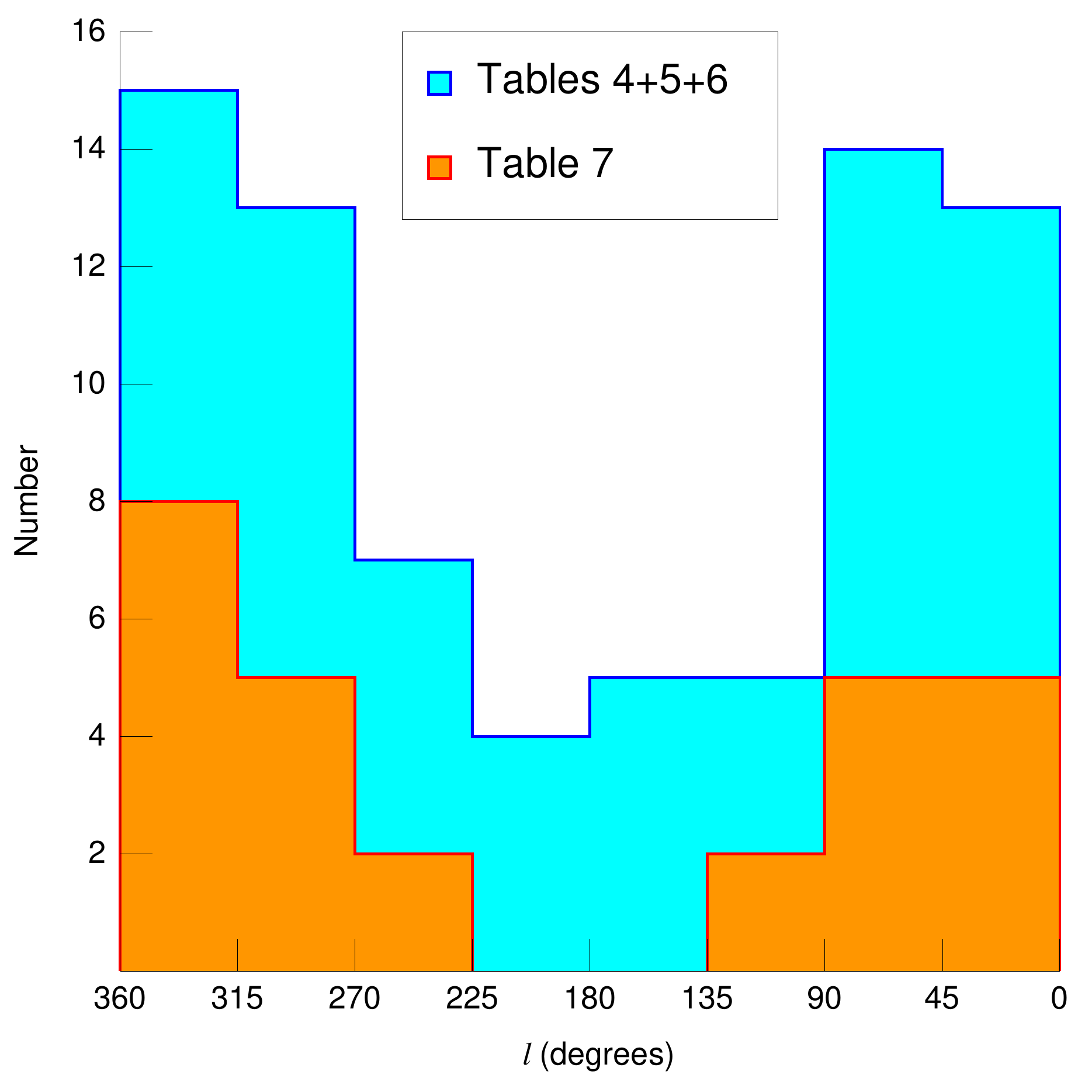}}
\caption{Galactic longitude histograms for O stars (blue/cyan) and supergiants (red/orange). The left panel shows the whole sample and the right panel the
runaway candidates.}
\label{lonhisto}
\end{figure*}

\subsection{Completeness}

\begin{table}
\caption{Robust parameter fits for the O stars and BA supergiants in this paper.}
\label{parameters}
\begin{center}
\begin{tabular}{lrrrrrr}
\hline
Sp. ty. & $<\!\mu_b\!>$ & $\sigma_{\mu_b}$ & $a_0$   & $a_1$   & $a_2$ & $\sigma_{\mu_l}$ \\
\hline
O       & $-$0.83       & 1.26             & $-3.04$ & $-2.07$ & 3.21  & 1.73             \\
BA I    & $-$0.88       & 1.03             & $-2.72$ & $-2.51$ & 3.35  & 1.54             \\
\hline
\end{tabular}
\end{center}
\end{table}

$\,\!$\indent To check for false negatives in our list, we searched \citet{Tetzetal11} for runaway candidates with 
a large probability of having a peculiar velocity\footnote{The peculiar velocity ${v_{\rm pec}}$ is the velocity of the object with respect to 
the average disk velocity at its location.} ($P_{v_{\rm pec}} >$ 0.5) that are missing % CHANGE
in Table~\ref{tabs1} but are present in our sample 1. There are 33 objects missing but, 
of those, 30 were detected by \citet{Tetzetal11} based mainly on their radial velocities as they have 
(a) $P_{v_{\rm r,pec}} > P_{v_{\rm t,pec}}$ and (b) $P_{v_{\rm t,pec}} < 0.5$. Therefore, we would not expect them 
to be detected by our 2-D method. Of the remaining three objects, one is HD~93\,521, which has no TGAS data and is
the highest - by far - latitude Galactic O star ($b = 52^{\rm o}$), something that makes it
difficult to detect in a 2-D method optimized for objects near the Galactic Plane. The other two, HD~108
and HDE~227\,018, have TGAS proper motions with significantly smaller uncertainties and closer to the
mean values than the Hipparcos values, which were the ones used by \citet{Tetzetal11}. Hence, a 3-D 
reanalysis would likely reduce their $P_{v_{\rm t,pec}}$. We conclude that our method 
correctly picks up those runaway stars with large tangential velocities but, as expected, misses some 
which are moving mostly in a radial direction. 
For O stars the 2-D method detects approximately half of the stars
a 3-D method would pick but the exact number is hard to evaluate given the differences between our sample 1 and that 
in \citet{Tetzetal11}.

What about false positives
(objects we identify as runaway stars but are not)? In their work on main-sequence B-type (i.e. lower mass) runaways, \citet{SilvNapi11}
describe their efforts to clean their sample of hot evolved low-mass stars, in their case Blue Horizontal Branch (BHB) stars. For our higher-mass
sample the expected fraction of such contaminants should be lower, as PAGB stars are the possible culprits (see Fig.~1 of \citealt{SilvNapi11})
and that phase is much shorter than the BHB one and only overlaps with our objects at the lower luminosity end. Another possible contaminant are 
sdO stars but those objects are excluded by the GOSSS spectral classification (see footnote 3). Therefore, we expect our false positives (if they exist)
to have other origins such as incorrect astrometric values. % CHANGE
The final answer about the number of false positives will lie, of course, in future work, but there is a good reason
why the new 13 objects in sample 2 had not been detected before as runaways. Eight of them do not have Hipparcos proper
motions and the remaining five were not included in \citet{Tetzetal11}. Therefore, we think that most of the runaway O candidates
will be confirmed in the future, possibly as soon as {\it Gaia} DR2 data become available.

A completeness analysis for BA supergiants is more difficult than for O stars for three reasons: 

\begin{itemize}
 \item There is currently no whole-sky spectroscopic survey of B supergiants analogous to GOSSS for O stars and the fraction of false identifications
       is high, as it is easy to find lower-luminosity B stars misclassified as supergiants. Hence, the sample is poorly defined and we are forced
       to reject a significant fraction of the potential runaways (Table~\ref{samples}) either because we have obtained a spectrum and found the 
       object was not a B supergiant or because there is no good-quality spectral classification in the literature.
 \item The situation is even worse for A supergiants. These are rare objects due to the speed at which massive stars cross this region of the HR diagram
       and there are just a few classification standards in the literature. 
 \item Most previous runaway studies have paid decreasing attention from stars of O type (through early-B and late-B) to A supergiants, making it
       difficult to compare our results with previous ones.
\end{itemize}

Therefore, we can only extrapolate from the O-star case to estimate that Table~\ref{tabs3b} will contain few false positives but
we cannot provide a number of expected false negatives.

\subsection{Comparing runaway O stars and runaway BA supergiants}

$\,\!$\indent In this subsection we compare our results for O stars and BA supergiants and analyze their similitudes and differences. In the first place,
we look at the fit parameters for the mean proper motions in Table~\ref{parameters} and Figs.~\ref{pmlat}~and~\ref{pmlon}. The fits are very similar and the
differences are hard to appreciate between the top and bottom panels in each figure. The dispersions for O stars are only slightly larger. This result
indicates that the bulk of both samples (i.e. the non-runaway stars) have similar properties in terms of distances and kinematics, as expected given that
during the lifetimes of BA supergiants they cannot travel far from their birth places (in a Galactic scale) unless they are ejected as runaways.

Next, we look at the distribution of stars by Galactic longitude in Fig.~\ref{lonhisto}. The left panel shows the distributions of the whole samples: O stars
and BA supergiants follow a similar pattern, with a marked concentration towards the inner Galaxy. The concentration is the result of the strong negative
radial gradient in the density of massive stars in the Galaxy, which overcomes the also negative gradient in extinction in the local neighborhood
\citep{MaizBarb18}. The over-density of O stars with respect to BA supergiants in the second and seventh octants is caused by the presence of Cygnus and
Carina, respectively, which are the two regions in the solar neighborhood with the largest concentrations of O stars. The opposite is seen in the third
octant mostly due to the presence of the low-extinction supergiant-rich Perseus arm. The right panel shows the distributions
of the runaway candidates. For O stars there appears to be little change but for BA supergiants the concentration towards the inner Galaxy becomes
significantly more marked: there are no runaway BA supergiants in the outer two octants and only two are seen in the third and sixth octants. Even though
numbers are relatively low, we believe this is partially an extinction effect: in the inner Galaxy extinction is significantly higher but also patchy and associated
with star-forming regions \citep{MaizBarb18}. Runaway BA supergiants have more time to travel away from the birth places and find locations where the
sightline is significantly less extinguished, especially if the ejection velocity vector has a significant vertical component.

\begin{figure}
\centerline{\includegraphics[width=\linewidth]{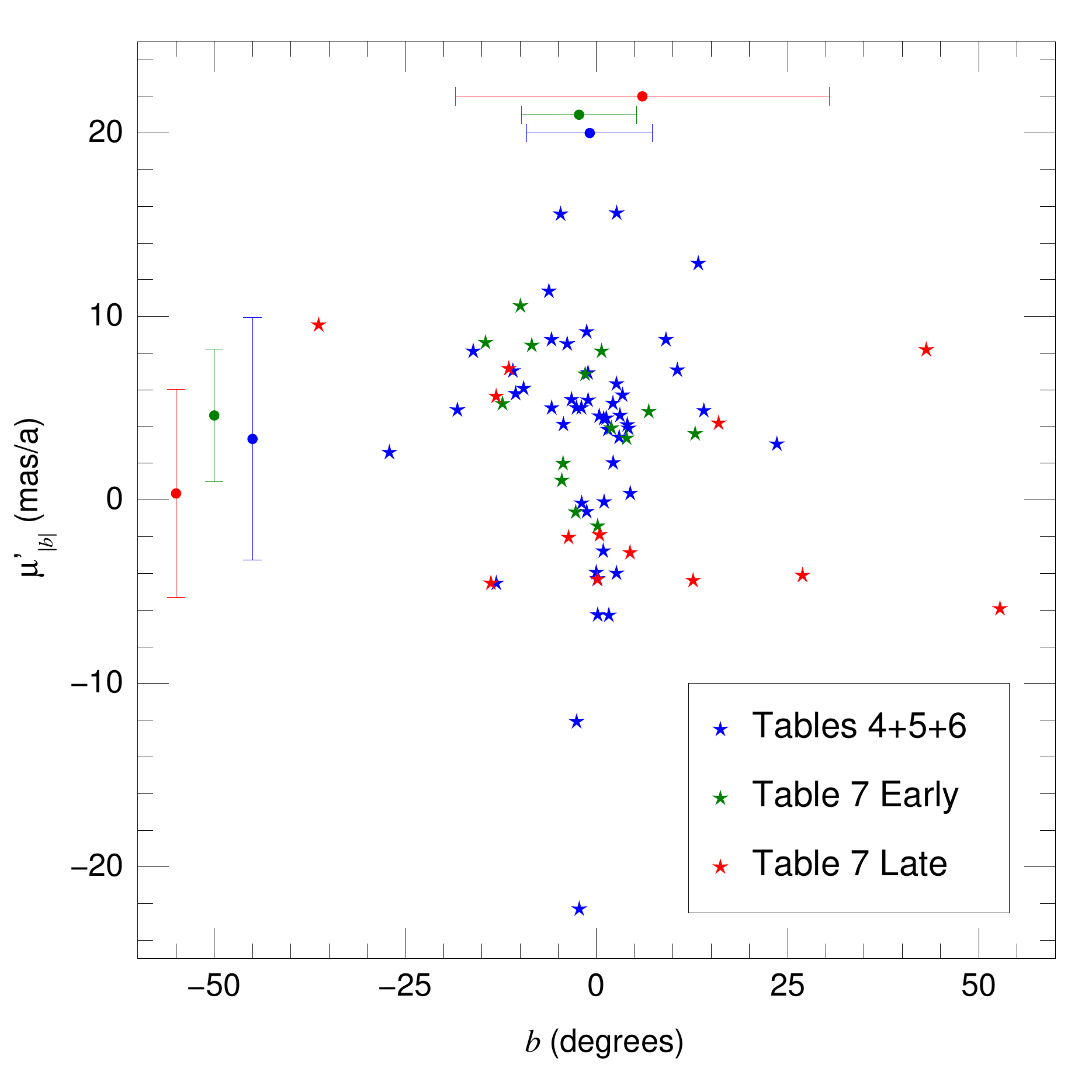}}
\caption{Corrected latitude proper motions in the absolute sense (positive for motions away from the Galactic Plane, negative for falling back motions)
as a function of Galactic latitude for runaway O stars (blue), B0-B0.7 supergiants (green), and B1 and 
later-type % CHANGE
supergiants (red). The error bars show the mean and standard distribution for each sample and coordinate.}
\label{latpmlatabs}
\end{figure}

In Fig.~\ref{latpmlatabs} we analyze the relationship for runaway stars between Galactic latitude and its corrected proper motion in the absolute 
sense, $\mu_{|b|}^\prime$, that is $\mu_b^\prime$ multiplied by the sign of $b$ so that the quantity is positive if the star is moving away from the Galactic
Plane and negative if it is falling back towards it. We have divided the BA supergiants into two subsamples: those with spectral subtypes B0-B0.7 and those
with later spectral subtypes. Regarding $b$, we see that runaway O stars and runaway early-B supergiants have similar dispersions and with a mean close to
zero (no preference for either Galactic hemisphere)\footnote{Remember that HD~93\,521 is not included in the sample, but the addition of a single star would
not change the result significantly.}. This indicates that there are no large time-difference effects between those two groups: an early-B supergiant is
older than a mid- or early-O dwarf but that is not necessarily true if the comparison is made with a late-O dwarf, which evolves into an early-B giant (not
supergiant). On the other hand, runaway 
later-type % CHANGE
supergiants have a significantly higher dispersion in Galactic latitude, indicating that they have been able 
to travel farther away from the Galactic Plane due to their longer average ages. 

With respect to $\mu_{|b|}^\prime$, we also see similitudes between runaway O stars and runaway early-B stars: both have average values that indicate that 
they are more likely to be moving away from the Plane than falling towards it. The dispersion is higher for O stars but this may be explained by the 
existence of some stars that are close to crossing the Galactic Plane for the first time due to a very recent ejection (two good examples of this are AE~Aur, 
\citealt{Hoogetal01}, and HD~155\,913). On the other hand, runaway 
later-type % CHANGE
supergiants have an average value of $\mu_{|b|}^\prime$ which is very close to zero:
this indicates that enough flight time has passed for about half of them to start falling back towards the Galactic Plane.

Finally, we consider whether there are differences in the fraction of O stars and BA supergiants that are runaways. A direct reading of Table~\ref{samples}
gives a value of 5.7\% fo the runaway fraction for O stars. However, as we have previously mentioned, a comparison with \citet{Tetzetal11} points towards an
incompleteness close to one half, mostly due to the use of a 2-D method for the detection of runaways. Therefore, a more realistic number would be in the
10-12\% range. Regarding BA supergiants, the number in Table~\ref{samples} is significantly lower than for O stars. There are at least three reasons why the
real fraction is likely to be different:

\begin{itemize}
 \item The use of a 2-D method makes us miss some runaways i.e. the same reason as for O stars.
 \item We have followed more strict criteria for the selection of BA supergiants, resulting in a higher fraction of discarded objects in Table~\ref{samples}.
       In some of those cases we have been able to confirm that the object was not a BA supergiant but in others we have not.
 \item As we have seen, runaway 
       later-type % CHANGE
       supergiants are located, on average, farther away from their birthplaces and farther away from the Galactic Plane with
       respect to the other runaways. That makes them easier to detect, as the main limitation to observe Galactic massive stars is extinction, not distance
       \citep{MaizBarb18}.
\end{itemize}

The first two reasons play in favor of increasing the real fraction of runaway BA supergiants while the last one against it. If we assume that the first
reason has a similar effect for supergiants than for O stars and that the last two cancel out (at least approximately), we are left with a real fraction of
runaway supergiants close to 6\% i.e. lower than for O stars. However, that value is just an estimate that needs to be confirmed by better data in the form
of a better sample selection and improved data.

\subsection{Runaways and rotational velocity}

\begin{table}
\caption{Line-width index statistics for the GOSSS I+II+III O-star sample and the O stars identified as runaways in this paper with GOSSS spectral 
classifications.}
\label{linewidth}
\begin{center}
\begin{tabular}{lrrrrrr}
\hline
Sample          & Total & Empty & (n)  & n  & nn \\
\hline
GOSSS I+II+III  & 590   & 435   & 106  & 40 &  9 \\
O runaways      &  43   &  23   &  14  &  5 &  1 \\
\hline
\end{tabular}
\end{center}
\end{table}

$\,\!\indent$ Runaways produced by supernova explosions are expected to have large rotational velocities \citep{Blaa93}, a characteristic that was verified by
\citet{Hoogetal01} for Galactic stars with a rather small sample and later on by \citet{Walbetal14} for 30 Doradus (see their Figure 9). 
Here we can use the homogeneous GOSSS sample of O stars to test whether runaway stars have larger rotational 
velocities compared to normal O stars. Each GOSSS spectral classification includes a line-width index that is empty for stars with low values of $v\,\sin i$
and is (n), n, or nn for stars with increasingly larger values of the projected rotational velocity\footnote{The approximate values for $v\,\sin i$ for 
(n), n, and nn are 200~km/s, 300~km/s, and 400~km/s, respectively, see \citet{Walbetal14}.}. More specifically, the line-width index measures the
broadening of intrinsically narrow lines (metallic absorption lines are preferred over He\,{\sc i} ones and those over He\,{\sc ii} lines) which is linked to
rotation in most cases. However, some stars with apparent broad lines are instead unresolved (in velocity) spectroscopic binaries. Over time, we have been
repeating GOSSS observations of stars with broad lines in order to resolve such cases and give independent spectral classifications to each component but a
minority of contaminants should still be present in the sample.

We present in Table~\ref{linewidth} the statistics on the line-width index for the 590 O stars in GOSSS I+II+III (the control group) and the 44 O stars 
identified as runaways in this paper with GOSSS spectral classifications. Note that most of the objects in the second group (along with the undetected O-type 
runaways) are included in the first group. For the control group 26.2$\pm$1.8\% (155/590) of the objects have non-empty line-width indices (fast rotators)
while for the O runaways the corresponding fraction is 46.5$\pm$7.6\% (20/43). Considering that we expect to be missing about half of the O-type runaways, the 
GOSSS sample is expected to contain close to 500 non-runaways of which $\sim$115 ($\sim$19\%) will have non-empty line-width indices (if the characteristics of 
the undetected runaways are similar to those of the detected ones). Therefore, we can conclude that Galactic runaway O-stars rotate significantly faster on 
average than their non-runaway counterparts. 

We also note that there is a significant difference between O supergiants (understood as luminosity classes II to Iab) and O stars in lower luminosity classes.
Among the former we find 3 fast rotators and 11 slow rotators while among the latter there are 17 fast rotators and 12 slow rotators. This is likely to be an 
age effect, as the rotational velocity is expected to decrease due to mass loss with a spin-down time scale of a few Ma \citep{Lauetal11}. This result points 
towards fast rotators being a majority of our runaway sample at ejection, though note that statistics may be biased in either direction: some fast rotators will 
not be identified as such due to their low value of $\sin i$ and some apparent fast rotators may actually be SB2s (e.g. HD~155\,913).
On the other hand, some runaways are the product of a multiple ejection (e.g. AE~Aur and $\mu$~Col, \citealt{Hoogetal00}) and others such as Y~Cyg 
\citep{Harmetal14} and AB~Cru (this paper) are close binaries. In both situations, the alternative scenario (a dynamical encounter in a compact stellar cluster) is
required to explain their origin. Therefore, it seems that both scenarios contribute to the population but more runaways 
in our sample appear to be produced in supernova explosions, though better statistics are needed to confirm this claim. On the other hand, the study of 
\citet{SilvNapi11} did not find such a prevalence of fast rotators but their sample was composed of main-sequence B stars i.e. of lower mass. Such a difference
could be a mass effect, as it should be easier to eject a 5~M$_\odot$ at a high velocity in a dynamical interaction than a 25~M$_\odot$ one. % CHANGE

\section{Summary}

The main results of this paper can be summarized as:

\begin{itemize}
 \item We present 76 runaway star candidates, of which 17 had no clear prior identification as such, 2 are unclear cases, and 13 were presented as 
       candidates for the first time in Paper I.
% T1 T2 T3 T4 SUM
% 29  0  2 15  46 (2 unclear)
%  0 13  0  0  13
%  0  0  5 12  17
%----------------
% 29 13  7 27  76
 \item Spectral classifications are assigned to 25 massive stars using GOSSS data and their spectrograms are included.
 \item We provide photometry-based $T_{\rm eff}$ estimates for five stars and present {\it WISE} imaging for twelve runaway candidates.
 \item The 2-D method used is estimated to detect approximately half of the runaways in the sample.
 \item We detect differences in the distribution of runaway B1 and 
       later-type % CHANGE
       supergiants compared to earlier runaways and we ascribe the differences to an age effect.
 \item Tentatively, the fraction of runaway BA supergiants appears to be lower than that of runaway O stars.
 \item Runaway O-star rotate on average faster that normal O stars with the supernova explosion scenario possibly contributing more objects to the population
       than the multi-body interaction one.
\end{itemize}

\begin{acknowledgements}
%{\bf We thank the referee, Ralf Napiwotzki, for helpful comments.} % CHANGE
This work makes use of 
[a] data from the European Space Agency (ESA) mission {\it Gaia} 
({\tt https://www.cosmos.esa.int/gaia}), processed by the {\it Gaia} Data Processing and Analysis Consortium (DPAC, 
{\tt https://www.cosmos.esa.int/web/gaia/dpac/consortium}) whose funding is provided by national institutions, 
in particular the institutions participating in the {\it Gaia} Multilateral Agreement;
 [b] data products from the Wide-field Infrared Survey Explorer ({\it WISE}), which is a joint project of the University of 
California, Los Angeles, USA, and the Jet Propulsion Laboratory/California Institute of Technology, USA,
funded by the National Aeronautics and Space Administration of the USA; and 
[c] the SIMBAD database and the VizieR catalogue access tool, both operated at CDS, Strasbourg, France.
J.M.A., A.S., and E.T.P. acknowledge support from the Spanish Government Ministerio de Econom{\'\i}a, Industria y Competitividad (MINECO/FEDER) through 
grant AYA2016-75\,931-C2-2-P. M.P.G. acknowledges support from the ESAC Trainee program.
R.H.B. acknowledges support from the ESAC Faculty Council Visitor Program.
S.S.-D., I.N., and E.T.P. acknowledge support from the Spanish Government Ministerio de Econom{\'\i}a, Industria y Competitividad (MINECO/FEDER) through 
grant AYA2015-68\,012-C2-1/2-P. 
\end{acknowledgements}

\bibliographystyle{aa}
\bibliography{general}

\end{document}